\documentclass[12pt]{article} 
\usepackage{amsfonts}
\usepackage{latexsym}
\usepackage{amsmath,amssymb}
\usepackage{verbatim}
\usepackage{mathtools}
\usepackage{color}
\usepackage[color,matrix,arrow]{xy}
\usepackage{changebar}
\usepackage{setspace}
\usepackage{simplewick}
\usepackage{tocvsec2}
\usepackage{young}
\usepackage[vcentermath]{youngtab}

\numberwithin{equation}{section}

\newcommand{\bea}{\begin{eqnarray}}
\newcommand{\eea}{\end{eqnarray}}
\newcommand{\be}{\begin{equation}}
\newcommand{\ee}{\end{equation}}
\newcommand{\ba}{\begin{align}}
\newcommand{\ea}{\end{align}}

  \makeatletter
  \let\over=\@@over \let\overwithdelims=\@@overwithdelims
  \let\atop=\@@atop \let\atopwithdelims=\@@atopwithdelims
  \let\above=\@@above \let\abovewithdelims=\@@abovewithdelims

\begin{document}

\ \\

\begin{center}

{\LARGE {\bf Perturbations of ${\cal W}_{\infty}$ CFTs}}

\vspace{0.8cm}

{\large Matthias R.\ Gaberdiel$^{a}$, Kewang Jin$^a$, and Wei Li$^b$}
\vspace{1cm}

{\it $^a$Institut f\"ur Theoretische Physik, ETH Zurich, \\
CH-8093 Z\"urich, Switzerland \\
\tt{\small gaberdiel@itp.phys.ethz.ch}, \tt{\small jinke@itp.phys.ethz.ch}
 }

\vspace{0.9cm}

{\it $^b$Max-Planck-Institut f\"ur Gravitationsphysik, Albert-Einstein-Institut, \\
Am M\"uhlenberg 1, 14476 Golm, Germany} \\
 \tt{\small wei.li@aei.mpg.de}

\vspace{1.0cm}

\end{center}

\begin{abstract}

The holographic duals of higher spin theories on AdS$_3$ are described by the large $N$ limit of 
a family of minimal model CFTs, whose symmetry algebra is equivalent to 
${\cal W}_{\infty}[\lambda]$. We study perturbations of these limit theories, and show that they
possess a marginal symmetry-preserving perturbation that describes switching on the $\frac{1}{N}$ 
corrections. We also test our general results for the specific cases of $\lambda=0,1$, where 
free field realisations are available.

\end{abstract}

\pagestyle{empty}

\pagebreak
\setcounter{page}{1}
\pagestyle{plain}

\setcounter{tocdepth}{2}
\begin{singlespace}
\tableofcontents
\end{singlespace}

\section{Introduction}

The proposed dualities relating a higher spin theory on AdS$_{d+1}$ to vector-like nearly free
conformal field theories in $d$ dimensions constitute simplified versions of the AdS/CFT
correspondence that are under very good quantitative control. As such they may open the way
towards understanding their inner workings.
The prototype example was proposed some years ago by
Klebanov \& Polyakov \cite{Klebanov:2002ja}, see also 
\cite{Sundborg:2000wp, Witten, Mikhailov:2002bp, Sezgin:2002rt} for earlier work
and \cite{Sezgin:2003pt} for a subsequent generalisation. It relates a Vasiliev higher spin theory
\cite{Vasiliev:1989re} on AdS$_4$ to the large $N$ limit of the ${\rm O}(N)$ vector model in $3$ 
dimensions. Compelling evidence for this duality was recently 
found by comparing correlation functions 
in \cite{Giombi:2009wh, Giombi:2010vg}, as well as through the work  
\cite{Maldacena:2011jn, Maldacena:2012sf} that determined interesting general constraints on the structure of 
the correlation functions based on symmetry considerations. More recently, these dualities were further
generalised to a one-parameter family of (in general) parity-breaking theories in 
\cite{Aharony:2011jz,Giombi:2011kc,Chang:2012kt}.

In a somewhat independent development, a lower dimension analogue of this duality was proposed in
\cite{Gaberdiel:2010pz}, relating a higher spin theory on AdS$_3$ 
\cite{Prokushkin:1998bq, Prokushkin:1998vn} to the large $N$ limit of 
a family of minimal model 2d CFTs. One of the guiding principles in proposing this 
duality was the observation that the asymptotic algebra of the higher spin theory on 
AdS$_3$ is described by a ${\cal W}_{\infty}[\lambda]$ algebra 
\cite{Henneaux:2010xg,Campoleoni:2010zq,Gaberdiel:2011wb,Campoleoni:2011hg}, which therefore largely
controls the dual CFT. This proposal was subsequently checked in various ways 
\cite{Gaberdiel:2011zw,Gaberdiel:2012ku}.

In the 3d/2d case, the underlying symmetry algebra ${\cal W}_{\infty}[\lambda]$ is characterised, in 
addition to the central charge (that corresponds geometrically to the size of the AdS space), by a continuous
parameter $\lambda$ that controls the higher spin interactions.\footnote{This parameter is the natural analogue of the 
$\lambda=\frac{N}{k}$ parameter in one dimension higher, where $N$ is the rank of the gauge group and $k$ the 
Chern-Simons coupling constant \cite{Aharony:2011jz,Giombi:2011kc,Chang:2012kt}.} From the dual minimal
model perspective, $\lambda$ is identified with
\begin{equation}
\lambda = \frac{N}{N+k} \ , \qquad \hbox{while} \quad
c = (N-1) \Bigl[ 1 - \frac{N(N+1)}{(N+k)(N+k+1)} \Bigr] \ ,
\end{equation}
where $k$ is the level of the coset model. 
In particular, $\lambda$ therefore becomes a continuous parameter in the 't~Hooft limit, in which
$c \sim N (1-\lambda^2)$. However, 
from the viewpoint of the ${\cal W}_{\infty}[\lambda]$ symmetry algebra, $\lambda$ and $c$ are arbitrary
(finite) parameters, and there is a priori no need to take $c \rightarrow \infty$ in order for $\lambda$
to become continuous.

It is then natural to ask whether theories corresponding to different values of $\lambda$ (and $c$) are 
connected by continuous deformations. For example, one may wonder whether there exist exactly
marginal operators that change $\lambda$ continuously (while preserving the ${\cal W}_{\infty}$ 
symmetry) without affecting $c$. Or there could be deformations that change both $\lambda$ and $c$
infinitesimally. Given that the ${\cal W}_{N,k}$ minimal models do not possess any exactly marginal 
operators, one may suspect that the first option is not possible, and indeed there is a fairly
general argument --- due to Stefan Fredenhagen --- that implies that such deformations cannot exist.
(This will be briefly reviewed in section~\ref{sec:structure}.) However, we find evidence that
a marginal deformation exists, at least in the 't~Hooft limit, that modifies both $\lambda$ and $c$.
In fact, it can be identified with a perturbation that introduces $\frac{1}{N}$ and $\frac{1}{k}$ corrections 
such that to first order $\lambda$ does not change.

More specifically, we characterise quite generally the perturbing operators that preserve the ${\cal W}_{\infty}$ 
symmetry to first order, and find that there is a one-parameter family of operators with this property.
They are uniquely characterised by their eigenvalues of the spin-$3$ generator. Quite remarkably,
the $h=1$ descendant of the light state corresponding to $(\tiny\yng(1);\tiny\yng(1))$ has
this property in the 't~Hooft limit and therefore defines (together with its conjugate operator) an
interesting perturbing field. We analyse its properties, first for the special cases $\lambda=0,1$, where
free field realisations are available,\footnote{These are the natural analogues of the free ${\rm O}(N)$ 
vector models in $3$ dimensions.}  and then for general $\lambda$, using conformal perturbation theory. 
The conformal perturbation theory is quite intricate since, in the 't~Hooft limit, there are infinitely many
fields of the same conformal dimension, and hence we need to analyse an infinitely degenerate case.
However, because of the structure of the fusion rules, the perturbation problem has a lot of structure,
and we can identify at least certain qualitative features. 
\smallskip

The paper is organised as follows. We begin with reviewing the free field realisations of the
special ${\cal W}_{\infty}[\lambda]$ theories at $\lambda=0$ (section~\ref{sec:mu0}) and $\lambda=1$ (section~\ref{sec:mu1}).
In particular, we show that the singlet sector of $N$ complex free fermions defines a ${\cal W}_{\infty}[0]$
algebra at $c=N-1$, while the singlet sector of $k$ complex free bosons lead to ${\cal W}_{\infty}[1]$
at $c=2k$. While both of these statements are certainly expected, they had not been established at
finite $c$ before; in particular, we confirm that the structure constants of the ${\cal W}_{\infty}[\lambda]$ algebras
agree precisely with the prediction of \cite{Gaberdiel:2012ku} for these values of $\lambda$ and $c$. 
In section~\ref{sec:deform} we then study the conditions a perturbing operator must satisfy in order
to preserve the ${\cal W}_{\infty}$ algebra at first order. This is first done in the 't~Hooft limit, and then
for finite $c$, using the full quantum version of the ${\cal W}_{\infty}[\lambda]$ algebra. 
We give strong evidence that the $h=1$ descendant of the light state 
corresponding to $(\tiny\yng(1);\tiny\yng(1))$ has this property in the 't~Hooft limit, and confirm this statement
explicitly for the special cases $\lambda=0$ and $\lambda=1$ (where the statement remains true
even at finite $c$). In section~\ref{sec:pert} we then study the effect of the perturbation by this field 
(and its conjugate) in the 't~Hooft limit, and identify the perturbed spectrum with what is obtained from
the minimal model perspective upon switching on a certain combination of $\frac{1}{N}$ and $\frac{1}{k}$ corrections.  
Finally, we conclude in section~\ref{sec:conc}. There are four appendices where some of the more 
technical material has been collected together.

\section{The theory at $\lambda=0$}\label{sec:mu0}

The simplest explicit realisation of the ${\cal W}_{\infty}[\lambda]$ algebra exists at $\lambda=0$, where we can 
describe the theory in terms of free fermions. We shall later also comment on how this description is related to
the continuous orbifold point of view proposed in \cite{Gaberdiel:2011aa}.

\subsection{The free fermion description}

Let us consider the theory of $N$ free complex fermions $\psi^i$ and $\psi^{*i}$, $i=1,\ldots, N$ with action 
\begin{equation}\label{feract}
S_0=\int d^2 z \left( \psi^{*i} \, \bar{\partial}\psi^i + \bar{\psi}^i \, \partial \bar{\psi}^{*i} \right) \ .
\end{equation}
The corresponding equations of motion are 
\begin{equation}\label{fereoms}
\bar{\partial}\psi^i=\bar{\partial}\psi^{*i}=0 \ , 
\end{equation}
and the OPEs take the form
\be
\psi^i(z_1) \, \psi^{\ast j}(z_2) \sim \frac{\delta^{ij}}{(z_1-z_2)} \ , 
\ee
with similar expressions for the right movers $\bar\psi^i$ and $\bar\psi^{*j}$. 
We shall always consider only states that are singlets with respect to the global ${\rm SU}(N)$ action. 

The free theory has the conserved spin-$s$ chiral currents \cite{Bergshoeff:1989ns,Bergshoeff:1990yd,Depireux:1990df}
(the expressions for the anti-chiral currents are analogous) 
\begin{equation}\label{Ws}
W^{s}=n(s)\sum^{s-1}_{k=0}(-1)^k \binom{s-1}{k}^2\,  \partial^{s-1-k}\psi^{*i}\,  \partial^{k}\psi^i  \ ,
\end{equation}
where the sum over $i$ is implicit, and we choose the normalisation convention 
\begin{equation}
n(s)=\frac{[(s-1)!]^2}{(2s-2)!} \ .
\end{equation}
Explicitly, for small values of $s$, these current are then
\begin{eqnarray}
J & \equiv & W^1 =  \psi^{*i}  \psi^i \\
 T & \equiv & W^2 = \frac{1}{2} (\partial\psi^{*i} \, \psi^i - \psi^{*i} \, \partial\psi^i )\\[4pt]
W & \equiv & W^3 = \frac{1}{6}\Bigl(\partial^2 \psi^{* i} \, \psi^i -4 \, \partial \psi^{* i} \, \partial\psi^i 
+ \psi^{* i} \,\partial^2\psi^i \Bigr)\\[4pt]
U & \equiv & W^4= \frac{1}{20} \Bigl( \partial^3\psi^{\ast  i} \, \psi^i - 9 \, \partial^2\psi^{\ast i} \, \partial \psi^i  
+ 9 \, \partial\psi^{\ast i} \, \partial^2 \psi^i   -  \psi^{\ast i} \, \partial^3 \psi^i   \Bigr) \ .
\end{eqnarray}

\subsection{OPEs and commutation relations}

It follows from the OPEs given in Appendix~\ref{app:OPE} that the modes of the 
stress energy tensor $T$ satisfy a Virasoro algebra with central charge $c=N$,
\be\label{Vir}
{}[L_m,L_n] = (m-n) L_{m+n} + \frac{N}{12} m \, (m^2-1) \, \delta_{m,-n} \ ,
\ee
while $J$ is a ${\rm U}(1)$-current whose modes satisfy 
\be
[J_m,J_n] = N\, m \, \delta_{m,-n} \ , \qquad [L_m,J_n] = - n \, J_{m+n} \ . 
\ee
It also follows from (\ref{JW}) and (\ref{TW}) that $W$ is neither ${\rm U}(1)$- nor Virasoro-primary. Indeed, converted 
into modes, these OPEs imply that the commutation relations take the form
\be
{}[J_m,W_n] = 2 m L_{m+n} \ , \qquad
[L_m,W_n] = (2m-n) W_{m+n} + \tfrac{1}{6} m (m^2-1) J_{m+n} \ . 
\ee
The OPE of $W$ with itself (\ref{WW}) then leads to 
\begin{eqnarray}\label{WWcomm}
{}[W_m,W_n] & = &  \tfrac{2N}{360}\, m  (m^2-1)  (m^2-4) \delta_{m,-n} 
+ \tfrac{1}{15} (m-n) (2m^2 + 2n^2 - mn - 8) L_{m+n} \nonumber \\[2pt] 
& & + 2 (m-n) U_{m+n}  \ .
\end{eqnarray}
Note that this algebra is linear, i.e.\ the commutators (\ref{WWcomm}) do not involve 
the normal ordered $\Lambda^{(4)}=:LL:$ term that generically appears in this commutator.

\subsection{The ${\rm U}(1)$ coset}\label{sec:res}

The free fermion theory does not directly describe ${\cal W}_{\infty}[\lambda]$ for any value of $\lambda$, since
it contains a spin $1$ current $J$, and hence leads to ${\cal W}_{1+\infty}$. Furthermore, in the above basis (in which
the generators are bilinears in the fermions, and hence the OPEs do not contain any non-linear term) the generators 
$W^s$ with $s\geq 2$ do not close among themselves, as follows for example from (\ref{TW}). There is, however, 
a different basis in which ${\cal W}_{\infty}[0]$ appears as a subalgebra of the free fermion theory. In this basis, 
the generators are not just bilinears in the free fermions, and as a consequence the algebra will turn out to be 
non-linear.

The basic idea for finding this basis is inspired by the coset construction, and in effect, the resulting
construction is what the coset by the ${\rm U}(1)$-current $J$ would amount to. We can recursively 
construct currents $\tilde{W}^s$ of spin $s=2,3,\ldots$ that are primary with respect to $J$. In terms of modes this 
then means that the modes $\tilde{W}^s_m$ commute with $J_n$. Because of the Jacobi identity, the same is then also true 
for the commutators of $\tilde{W}^{s_1}$ and $\tilde{W}^{s_2}$, i.e.\ the $\tilde{W}^s$ generators form a closed algebra. In the following we shall construct the first few of these generators explicitly; we can then determine
their commutation relations, and show that they generate indeed ${\cal W}_{\infty}[0]$ at $c=N-1$.

\subsubsection{The coset generators}

Let us first define the generators $\tilde{W}_s$ recursively. For $s=2$, we get
\be\label{Tt}
\tilde{T} = T - \frac{1}{2N} : JJ: \ , 
\ee
or in terms of states 
\be \label{tildeL}
\tilde{L}_{-2} \Omega = L_{-2} \Omega- \frac{1}{2N} J_{-1} J_{-1} \Omega \ .
\ee
It is easy to see that (\ref{tildeL}) is ${\rm U}(1)$-primary, i.e.\   $J_n \tilde{L}_{-2}\Omega=0$ for $n\geq 0$. The OPEs
of $\tilde{T}$ with $J$ and itself are then
\be
\tilde{T} (z_1) J(z_2) \sim 0 \ , \qquad
\tilde{T} (z_1) \tilde{T}(z_2) \sim \frac{(N-1)/2}{(z_1-z_2)^4} + \frac{2 \tilde{T}(z_2)}{(z_1-z_2)^2} + \frac{\tilde{T}'(z_2)}{(z_1-z_2)} \ ,
\ee
i.e.\ the new modes $\tilde{L}_n$ define a Virasoro algebra (\ref{Vir}) with central charge $N-1$ (instead of $N$),
and commute with the modes $J_n$. 

\smallskip

\noindent 
For $s=3$, the ${\rm U}(1)$-primary generator is
\be\label{Wtf}
\tilde{W} = W - \frac{2}{N} : JT: + \frac{2}{3N^2}\, :JJJ: \ ,
\ee
or in terms of states
\be\label{Wt}
\tilde{W}_{-3} \Omega = W_{-3} \Omega - \frac{2}{N} \, J_{-1} L_{-2} \Omega + \frac{2}{3N^2} \, J_{-1} J_{-1} J_{-1} \Omega \ .
\ee
Moreover, $\tilde{W}$ is also primary with respect to $\tilde{T}$, i.e.
\be\label{TtW}
\tilde{T}(z_1) \tilde{W}(z_2) \sim \frac{3 \tilde{W}(z_2)}{(z_1-z_2)^2} + \frac{\tilde{W}'(z_2)}{(z_1-z_2)} \ ,
\ee
which in particular does not involve the current $J$ any longer, in contrast to  the situation in (\ref{TW}), and in 
agreement with the above general argument. 
Incidentally, the simplest way to compute (\ref{TtW}) is to use the general formula
\be\label{general}
\tilde{T}(z_1) \tilde{W}(z_2) = \sum_{n\in\mathbb{Z}} \, (z_1-z_2)^{-n-2} V\bigl( \tilde{L}_n \tilde{W}_{-3}\Omega,z_2 \bigr) \ , 
\ee
where $V(\psi,z)$ denotes the vertex operator associated to $\psi$, together  with the mode expansion of $\tilde{T}$,
\be
\tilde{L}_n = L_n - \frac{1}{2N} \sum_{l\in\mathbb{Z}} : J_{n-l} J_l: \ .
\ee
We also need the explicit formula for the ${\rm U}(1)$-primary state at $h=4$, which equals
\begin{eqnarray}
\tilde{U}_{-4} \Omega & = & U_{-4} \Omega - \frac{1}{5N} J_{-1} J_{-3} \Omega - \frac{3}{20N} J_{-2}J_{-2} \Omega
- \frac{3}{N} J_{-1} W_{-3} \Omega + \frac{3}{N^2} J_{-1} J_{-1} L_{-2}\Omega \nonumber \\
& & - \frac{3}{4N^3} J_{-1}J_{-1}J_{-1}J_{-1} \Omega - \frac{(21 
- \frac{15}{N})}{(5N+17)} \, (\tilde{L}_{-2} \tilde{L}_{-2} - \tfrac{3}{5} \tilde{L}_{-4}) \Omega \ , 
\end{eqnarray}
where the last term is required in order to make it also Virasoro primary with respect to $\tilde{T}$. 

Continuing in this manner, we can recursively construct ${\rm U}(1)$-primary fields $\tilde{W}^s$ that 
generate a closed algebra. We can furthermore recursively make them  Virasoro primary 
(with respect to $\tilde{T}$), and thus the resulting ${\cal W}_{\infty}$ algebra has the spin content $2,3,4,\ldots$. 
Following the general logic of \cite{Gaberdiel:2012ku}, it must therefore be isomorphic to ${\cal W}_{\infty}[\lambda]$
 for some value of $\lambda$. In the following we shall show that the relevant value of $\lambda$ is $\lambda=0$. Note that
 the classical analysis of \cite{Gaberdiel:2011wb} only tells us that this has to be true to leading order in 
 $1/c$; now we have shown that it is actually true even at finite $c=N-1$.

\subsubsection{Determining $\lambda$}

In order to determine the value of the parameter $\lambda$, it is sufficient to calculate two commutators. Indeed, 
in the conventions of \cite{Gaberdiel:2012ku} and \cite{Gaberdiel:2012yb}, we have\footnote{In 
\cite{Gaberdiel:2012yb} $N_4$ was defined with the opposite sign compared to \cite{Gaberdiel:2012ku}, as follows
from comparing the equation for the commutator $[W_m,U_n]$. Here we use the conventions of \cite{Gaberdiel:2012ku}.}
\begin{eqnarray}
{}[\tilde{W}_m,\tilde{W}_n] & = &  \frac{N_3 c}{144}\, m  (m^2-1)  (m^2-4) \delta_{m,-n} 
+ \frac{N_3}{12} (m-n) (2m^2 + 2n^2 - mn - 8) \tilde{L}_{m+n} \nonumber \\[2pt] 
& & + 2 (m-n) \tilde{U}_{m+n} + \frac{40\, N_3}{(5 c+22)} (m-n) \tilde{\Lambda}_{m+n} \ , \label{tWtWcomm} \\[4pt]
{}[\tilde{U}_m,\tilde{U}_n] & = & \frac{N_4 c}{4320} \, m  (m^2-1)  (m^2-4) (m^2-9) \delta_{m,-n} + \cdots \ , \label{UtUtcomm}
\end{eqnarray}
and the parameter $\gamma^2$, which determines the coefficient of $U$ in the $WW$ OPE
\cite{Gaberdiel:2012ku}  and  characterises the ${\cal W}$ algebra uniquely, is then given by
\be\label{gammadef}
\gamma^2 = \frac{896}{75} \, \frac{N_4}{N_3^2} \ . 
\ee
Using the analogue of (\ref{general}), we have worked out the first few terms of the $\tilde{W} \tilde{W}$ OPE to be
\begin{eqnarray}
\tilde{W}(z_1)\, \tilde{W}(z_2) & \sim  & 
\frac{2 (N-1)(N-2)}{3N } \, \frac{1}{(z_1-z_2)^6}  +  \frac{4(N-2)}{N} \frac{\tilde{T}(z_2)}{(z_1-z_2)^4}
+ \frac{2(N-2)}{N} \frac{\tilde{T}'(z_2)}{(z_1-z_2)^3} \nonumber \\[4pt]
& & + \frac{4\, \tilde{U}(z_2) + \frac{3}{5}\, \tilde{T}''(z_2)}{(z_1-z_2)^2} 
+ \frac{64 (N-2)}{5 N (N-1)}  \frac{\tilde\Lambda^{(4)}(z_2)}{(z_1-z_2)^2} + {\cal O}\left( \frac{1}{z_1-z_2} \right) \ , \label{WtWt}
\end{eqnarray}
where $\tilde{\Lambda}^{(4)}$ is the composite field $\tilde{\Lambda}^{(4)} = : \tilde{T} \tilde{T}: - \frac{3}{10}  \partial^2 \tilde{T}$. 
Comparing with the central term in (\ref{tWtWcomm}) and using that $c=N-1$, we conclude that 
\be
N_3 = \frac{4}{5} \frac{N-2}{N} \ . 
\ee
In order to compute $N_4$, we have also determined the central term in the $\tilde{U} \tilde{U}$ OPE, i.e.
\be
\tilde{U}_4 \tilde{U}_{-4}\Omega = \frac{9 (N-1)(N-2) (N-3)(N+1)}{N^2 (5 N+17)}\, \Omega  \ , 
\ee
which gives
\be
N_4 = \frac{54}{7}\, \frac{ (N-2) (N-3)(N+1)}{N^2 (5 N+17)}\ .
\ee
Thus $\gamma^2$ becomes
\be\label{gammaex}
\gamma^2 = \frac{ 144 (N+1)(N-3) }{(5N+17)(N-2)} = 
\left. \frac{64 (c+2) (\lambda-3) \bigl(c(\lambda+3)+2(4\lambda+3)(\lambda-1)\bigr)}{(5c+22)(\lambda-2)
\bigl(c(\lambda+2)+(3\lambda+2)(\lambda-1)\bigr)} \right|_{\lambda=0,c=N-1} \ ,
\ee
i.e.\ it agrees with the general formula of \cite[eq.~(2.15)]{Gaberdiel:2012ku} at $\lambda=0$ and $c=N-1$. This proves that
the free fermion construction indeed gives rise to ${\cal W}_{\infty}[\lambda=0]$ at $c=N-1$. 

We should mention that this argument relies on the assumption that the only consistent ${\cal W}_{\infty}$
algebras that are generated by one field of each integer spin $s\geq 2$ are described by ${\cal W}_{\infty}[\lambda]$. 
While this statement has not been established in complete generality, there is very convincing evidence, based on
the analysis of \cite{MC}, that this is indeed the case.\footnote{In \cite{MC} the commutators of the spin fields
with total spin $s\leq 10$ were determined uniquely in terms of $\lambda$ and $c$, using recursively
the Jacobi identities.}

\subsection{The continuous orbifold viewpoint}

It was argued in \cite{Gaberdiel:2011aa} that the theory at $\lambda=0$ can be described in terms of a continuous
orbifold. More specifically, one considers the affine $\hat{\mathfrak{su}}(N)_1$ theory, and takes the orbifold 
by the action of the group ${\rm SU}(N)/\mathbb{Z}_N$. In the untwisted sector this amounts to restricting the
affine level $1$ theory to those states that are ${\rm SU}(N)/\mathbb{Z}_N$ singlets. From this point of view, the higher 
spin currents then arise from the Casimir operators; in particular, the stress 
energy tensor of the continuous orbifold theory equals
\be\label{Tco}
T^{{\rm co}} = \frac{1}{2(N+1)}\, \sum_{a} :J^a J^a: \ ,
\ee
where $J^a$ are the currents of the $\hat{\mathfrak{su}}(N)_1$ theory, while the spin $3$ generator is
\be\label{Wco}
W^{\rm co} = \frac{1}{3(N+1)(N+2)}\,\sum_{abc} d_{abc} :J^a J^b J^c: \ , 
\ee
where $d_{abc}$ is the totally symmetric invariant tensor. The higher spin generators are similarly associated to
the higher order Casimir operators.

Actually, the relation between this continuous orbifold description and the free fermion construction from above 
is fairly immediate. The theory of $N$ complex fermions has a $\hat{\mathfrak{u}}(N)_1$ algebra, whose generators are 
\be
J^{ij} = \psi^{\ast \, i} \psi^j \ .
\ee
This current algebra contains the $\hat{\mathfrak{u}}(1)$ subalgebra generated by $J=\sum_i \psi^{\ast \, i} \psi^i$ and
the coset by this $\hat{\mathfrak{u}}(1)$ algebra (see section~\ref{sec:res}) leads to an affine $\hat{\mathfrak{su}}(N)_1$
theory at $c=N-1$.  Indeed, the $\hat{\mathfrak{su}}(N)_1$ currents are simply given by, see \cite[chapter 15.5.6]{DiFrancesco}
\be\label{Jiden}
J^a = \sum_{ij} \psi^{\ast i} \, t^a_{ij} \, \psi^j  \ ,
\ee
where $t^a_{ij}$ are the generators of $\mathfrak{su}(N)$ in the fundamental representation. In both theories
we are furthermore considering only singlets with respect to the global ${\rm SU}(N)$ action, and thus the spectra
agree precisely (in the untwisted sector). The twisted sectors are then completed by consistency, and thus should also match. 
(It might be interesting to understand the twisted sectors more directly from the free fermion point of view; this could 
be closely related to the discussion of \cite{Banerjee:2012gh} in one dimension higher.)

Using the translation between the free fermion and the $\hat{\mathfrak{su}}(N)_1$ description, it was shown in 
\cite[chapter 15.5.6]{DiFrancesco} that the stress energy tensor of (\ref{Tco}) agrees indeed with $\tilde{T}$ from 
(\ref{Tt}). We have also checked that (\ref{Wco}) and (\ref{Wtf}) similarly agree.

\section{The theory at $\lambda=1$}\label{sec:mu1}

The other simple realisation of the ${\cal W}_{\infty}[\lambda]$ algebra arises for $\lambda=1$, for which 
there is a description in terms of $k$ complex free bosons.

\subsection{The higher spin currents}

The theory of $k$ complex free bosons has spin-$s$ currents \cite{Bakas:1990ry}
\begin{equation}\label{Wbos}
W^s(z) = m(s) \sum_{l=1}^{s-1}  \, \frac{(-1)^l}{(s-1)} \left( {s-1 \atop l} \right) \left( {s-1 \atop s-l} \right) \,  \partial^l \phi^{\, j} \,\, \partial^{s-l} \bar{\phi}^{\, j}  \ , 
\end{equation}
where the sum over $j=1,\ldots, k$ is implicit, and we choose the normalization factor
\begin{equation}
m(s) = 
\frac{2^{s-3} s!}{(2s-3)!!}\ .
\end{equation}
Explicitly, the first few currents are 
\begin{eqnarray}
T(z) & \equiv & W^2(z) = - : \partial \phi^{\, j} \, \partial \bar{\phi}^{\, j} : \\[2pt]
W(z) & \equiv & W^3(z)  = -2 
: (\partial \phi^{\, j} \, \partial^2 \bar{\phi}^{\, j} - \partial^2 \phi^{\, j} \, \partial \bar{\phi}^{\, j} ) : \\[2pt]
U(z) & \equiv & W^4(z)  = - \frac{16}
{5} : ( \partial \phi^{\, j} \, \partial^3 \bar{\phi}^{\, j}
- 3 \, \partial^2 \phi^{\, j} \, \partial^2 \bar{\phi}^{\, j} + \partial^3 \phi^{\, j} \, \partial \bar{\phi}^{\, j} ) : \ .
\end{eqnarray}
Using the OPEs of the currents 
\begin{equation}
\partial \phi^{\, i} (z_1) \, \partial \bar{\phi}^{\, j} (z_2) \sim - \frac{\delta^{ij}}{(z_1-z_2)^2}
\end{equation}
we can work out the OPEs of higher spin currents, and one finds for the stress energy tensor
\begin{eqnarray}
T(z_1) T(z_2) = \frac{k}{(z_1-z_2)^4} + \frac{2 T(z_2)}{(z_1-z_2)^2} + \frac{\partial T(z_2)}{(z_1 - z_2)} + \cdots \ ,
\end{eqnarray}
thus showing that the central charge equals $c=2k$, as expected. Some of the other OPEs are worked
out explicitly in Appendix~\ref{app:bos}, see eqs.~(\ref{bos23}) -- (\ref{bos44}); converted into modes,
they give rise to the commutation relations 
\begin{eqnarray}
[L_m,L_n] &=& (m-n) L_{m+n} + \tfrac{k}{6} m(m^2-1) \delta_{m+n} \\[2pt]
[L_m,W_n] &=& (2 m-n) W_{m+n} \label{bosLW} \\[2pt]
[L_m,U_n] &=& (3 m-n) U_{m+n} +  \tfrac{32}{5} 
m(m^2-1) L_{m+n}  \label{bosLU} \\[2pt]
[W_m,W_n] &=& 2(m-n) U_{m+n} +  \tfrac{4}{5} 
(m-n)(2m^2+2n^2-mn-8) L_{m+n} \cr
&& +  \tfrac{2k}{15} 
m(m^2-1)(m^2-4) \delta_{m+n} \\[2pt]
[W_m,U_n] &=& (3m-2n) X_{m+n} +  \tfrac{64}{35} 
(5 m^3-5 m^2 n+3 m n^2-17 m-n^3+9 n) W_{m+n} \nonumber \\[2pt]
[U_m,U_n] &=& \tfrac{64}{525} 
k m(m^2-1)(m^2-4)(m^2-9) \delta_{m+n} + \cdots \ ,
\end{eqnarray}
where we have only worked out the leading term for the $[U,U]$ commutator.

\subsection{The primary basis}

The $W^s$ as defined in (\ref{Wbos}) are not primaries: while $W^{3}$ is still a primary (see (\ref{bosLW})), 
the spin-$4$ field $U$ is already not primary (see eq.~(\ref{bosLU})). 
In order to identify the $\lambda$ value of the resulting algebra, it is convenient to go to a primary basis. 
For instance, the spin-$4$ current in the primary basis is 
\be
\tilde{U}= U - \frac{192} 
{10k+22} \, \Lambda^{(4)} \ , 
\ee
where $\Lambda^{(4)} \left( = : T T : - \frac{3}{10} \partial^2 T \right)$ is the familiar composite field, see eq.~(\ref{WtWt}). 
In terms of this field, the relevant OPEs then become 
\begin{eqnarray}
W(z_1) W(z_2) &=&  \frac{16 k }
{z_{12}^6} +  48 
\left[ \frac{T}{z_{12}^4} + \frac{1}{2}\frac{\partial T}{z_{12}^3} 
+ \frac{3}{20}\frac{\partial^2 T}{z_{12}^2} + \frac{1}{30}\frac{\partial^3 T}{z_{12}} \right] \nonumber \\[2pt]
&& + \frac{4}{z_{12}^2} \left[ \tilde{U} + \frac{192} 
{10k+22} \Lambda^{(4)} \right]
+ \frac{2}{z_{12}} \left[ \partial \tilde{U} + \frac{192} 
{10k+22} \partial \Lambda^{(4)}  \right]  \ , \label{33OPE1} \\[8pt]
\tilde{U}(z_1) \tilde{U}(z_2) &=& \frac{3072 (2k+2) k}{10k+22} \frac{1 }
{z_{12}^8} + \cdots  \ .\label{44OPE1}
\end{eqnarray}
Comparing as before with eqs.~(\ref{tWtWcomm}) and (\ref{UtUtcomm}) we thus conclude that 
\be\label{3.16}
N_3 = \frac{48}{5} \ , \qquad N_4 =  \frac{9216}{7} \, \frac{c+2}{5c + 22} \ ,
\ee
where we have used that $c=2k$. Thus the $\gamma^2$ parameter from eq.~(\ref{gammadef}) becomes
\be
\gamma^2 = \frac{ 512 (c+2) }{3 (5c+22)} = 
\left. \frac{64 (c+2) (\lambda-3) (c(\lambda+3)+2(4\lambda+3)(\lambda-1))}{(5c+22)
(\lambda-2)(c(\lambda+2)+(3\lambda+2)(\lambda-1))} \right|_{\lambda=1} \ .
\ee
Hence the free boson theory generates indeed ${\cal W}_{\infty}[\lambda=1]$ at $c=2k$. Moreover, the seemingly non-linear
${\cal W}_{\infty}[\lambda=1]$ algebra of eq.~(\ref{33OPE1}) is in fact linear upon going to the original non-primary basis
(\ref{Wbos}), in agreement with the analysis of \cite{Gaberdiel:2011wb}.

\section{Deforming the theory}\label{sec:deform}

We are interested in perturbing the ${\cal W}_{\infty}[\lambda]$ algebra by a marginal operator that preserves
the current symmetry, i.e.\ by a perturbation that leaves all currents of ${\cal W}_{\infty}[\lambda]$ holomorphic. From 
the analysis of \cite{Fredenhagen:2007rx}, see also \cite{Cardy}, we know that to first order in perturbation 
theory this will be the case provided that 
\be
\lim_{\epsilon\rightarrow 0} \oint_{|w-z|=\epsilon} dw\, \Phi(w,\bar{w}) W^s(z) = 0 \ .
\ee
Since we know that the OPE of $W^s$ with $\Phi$ is of the form
\be
W^s(z)  \Phi(w,\bar{w})  = \sum_{l} (W^s_l \Phi)(w,\bar{w}) \, (z-w)^{-l-s}
\ee
the requirement that $W^s$ remains holomorphic means that 
\be\label{cond}
{\cal N}_{s} \equiv \sum_{l=0}^{s-1} \frac{(-1)^l}{l!}\, (L_{-1})^l\, W^s_{-s+1+l}  \, \Phi = 0 \qquad s\geq 2 \ .
\ee
(Here we have assumed that $\Phi$ is ${\cal W}_{\infty}[\lambda]$ primary.)
Given that the ${\cal W}_{N,k}$ minimal models are not expected to have any such
perturbation --- the integrable perturbation by the field $(0;{\rm adj})$ is relevant and induces the 
RG flow from $k\rightarrow k-1$ --- we can only hope to find a solution to this 
problem either at generic values of $(\lambda,c)$ of the ${\cal W}_{\infty}[\lambda]$ theory, 
or in the 't~Hooft limit of the ${\cal W}_{N,k}$ models. Remarkably, there is a simple perturbing field 
$\Phi$ that seems to satisfy (\ref{cond}) in the 't~Hooft limit, as we shall now explain.

\subsection{The perturbation in the 't~Hooft limit}\label{sec:perttH}

We begin by analysing the condition (\ref{cond}) in the 't~Hooft limit, i.e.\ in the limit of the minimal
models ${\cal W}_{N,k}$ where we take $N,k\rightarrow \infty$ while keeping
\be\label{lambdadef}
\lambda =  \frac{N}{N+k} 
\ee
fixed. It was shown in \cite{Gaberdiel:2012ku} that with the central charge 
\be\label{cNk}
c_{N,k} = (N-1) \Bigl[ 1 - \frac{N(N+1)}{(N+k)(N+k+1)} \Bigr] 
\ee
the chiral algebra ${\cal W}_{N,k} \cong {\cal W}_{\infty}[\lambda]$, even at finite $N$ and $k$. In the 't~Hooft 
limit $c_{N,k}\rightarrow \infty$, and hence the non-linear terms in ${\cal W}_{\infty}[\lambda]$ drop out. 
Let us denote the eigenvalues of the ${\cal W}_{\infty}[\lambda]$ 
primary state $\Phi$ by 
\be
W^t_0 \Phi = w^t \Phi \qquad t\geq 2 \ .
\ee
By construction ${\cal N}_2=0$ provided that $\Phi$ is marginal, i.e.\ $w^2 \equiv h =1$. 
On the other hand, the ${\cal N}_s$ with
$s\geq 3$ are generically non-trivial null-vectors. In order to confirm that they are indeed null we need
to show that they are annihilated by all positive ${\cal W}_{\infty}[\lambda]$ modes. Using the commutation relations
$[L_m,W^s_n] = ((s-1)m - n)  \, W^s_{m+n}$ one can easily check that 
\be
L_n\,  {\cal N}_s = 0 \qquad n\geq 1 \ , \ s\geq 3 \ .
\ee
Thus it is sufficient to require
\be\label{cond1}
W^s_{n} \, {\cal N}_t = 0 \ , \qquad s,t \geq 3\ , \  \ n\geq 1 \ .
\ee
We now consider the special case of (\ref{cond1}) with $s=3$ and $n=t-1$. Then using the 
structure of the commutation relations of ${\cal W}_{\infty}[\lambda]$, in particular the fact that
the highest spin mode that appears in the commutator of $W^3$ with $W^t$ is $W^{t+1}$, (\ref{cond1})
leads to an equation for the eigenvalue $w^{t+1}$ in terms of the eigenvalues $w^s$ with $s\leq t$
(as well as the structure constants of the algebra). For example, from (\ref{cond1}) with
$s=3$ and $n=t-1=2$ with $t=3,4,5$ we obtain 
\begin{eqnarray}
w^4 & = & \frac{3}{4} \Bigl(N_3  + (w^3)^2 \Bigr) \label{w41} \\[2pt]
w^5 & = & \frac{1}{15} \Bigl( 10\, w^4\, w^3 + 8 \, \frac{N_4}{N_3}\, w^3 \Bigr) \label{w51}  \\[2pt]
w^6 & = &  \frac{1}{9} \Bigl( 5 \, (w^4)^2 + 30 \, n_{44} \, w^4 - 3 N_4 \Bigr)  \ ,  \label{w61} 
\end{eqnarray}
where $N_3$, $N_4$ and $n_{44}$ are the structure constants of the ${\cal W}_{\infty}[\lambda]$
algebra, which equal, in our conventions, 
\be\label{structure}
N_3 = \frac{16}{5}\, \sigma^2 (\lambda^2-4) \ , \quad
N_4 = \frac{384}{35}\, \sigma^4 (\lambda^2-4)(\lambda^2-9) \ , \quad
n_{44} = \frac{8}{15}\, \sigma^2\, (\lambda^2 - 19) \ ,
\ee
where $\sigma$ is a normalisation constant.

Thus this subset of conditions fixes the eigenvalues of $\Phi$ up to an arbitrary choice of $w^3$. One might
wonder whether some of the other conditions in (\ref{cond1}) would then fix $w^3$, but this does not seem to be the 
case.\footnote{We have worked out a few more relations obtained from (\ref{cond1}), but they all follow from those
above.}
In fact there is an intuitive reason why this should be so: the set of conditions (\ref{cond1}) with $s=3$ and $n=t-1$ is 
equivalent to the set of conditions with $t=3$, $n=1,2$ and $s=3,4,\ldots$. On the other hand, these 
latter conditions are equivalent to the statement that ${\cal N}_3$ is indeed a null-vector, i.e.\ to the
statement that the $W^3$-current remains holomorphic to first order in perturbation. But since the whole ${\cal W}_{\infty}[\lambda]$
algebra is generated by repeated OPEs from $W^3$, this should then imply that the whole
${\cal W}_{\infty}[\lambda]$ algebra remains holomorphic to first order, i.e.\ (\ref{cond1}) is satisfied for all $s,t\geq 3$ and $n\geq 1$. 

This suggests that a one-parameter family of representations preserve the ${\cal W}_{\infty}[\lambda]$ algebra to first
order in perturbation theory. One might then wonder whether one of the coset fields would satisfy these constraints in the 't~Hooft limit. Quite remarkably, this does seem to be the case. To see how this goes, recall that the $(\tiny{\yng(1)};\tiny{\yng(1)})$
representation becomes indecomposable in the 't~Hooft limit, i.e.\ its structure is  \cite{Gaberdiel:2011zw}
\be\label{ffdia}
 ({\tiny{\yng(1)}};{\tiny{\yng(1)}}): \qquad 
\xymatrix@C=1pc@R=2.2pc{
   & \vdots & & \vdots & & \vdots & & \vdots \\
   & 2 & & \ar@<0.4ex>[dr]^{L_1} \rho & & \ar@<-0.4ex>[dl] \xi & & \ar@<-0.4ex>[lldd]_{L_2} T \\
  & 1 & & & \phi\ar@<0.4ex>[ul]^{L_{-1}} \ar@<-0.4ex> [ur] \ar[dr]_{L_1}& & &\\
 L_0 =& 0 & &  & & \omega\ar@<-0.4ex>[uurr]_{L_{-2}} & & 
}
\ee 
Furthermore, in a suitable rescaling limit (see Section~\ref{sec:pert}), the arrow between $\phi$ and $\omega$
disappears, and $\phi$ becomes a primary field of conformal dimension $h=1$. Its eigenvalues are simply
\begin{eqnarray}
w^s(\phi) = w^s ( \tiny{\yng(1)} ;0) + w^s(0;\tiny{\yng(1)}) \ ,
\end{eqnarray}
where $w^s(\tiny{\yng(1)};0)$ and $w^s(0;\tiny{\yng(1)})$ are the eigenvalues of $W^s_0$ on the highest weight states of $(\tiny{\yng(1)};0)$ 
and $(0;\tiny{\yng(1)})$, respectively. In the conventions of \cite{Gaberdiel:2012yb} the eigenvalues of these 
representations are
\begin{equation}
\begin{array}{rclrcl}
w^2(\tiny{\yng(1)};0) & = & \frac{1}{2} (1+\lambda) \qquad 
& w^2(0;\tiny{\yng(1)}) & = & \frac{1}{2} (1-\lambda) \\
w^3(\tiny{\yng(1)};0) & = & \frac{2}{3} \, i \, \sigma\, (1+\lambda) (2+\lambda) \qquad 
& w^3(0;\tiny{\yng(1)}) & = & - \frac{2}{3}\, i \,  \sigma\, (1-\lambda)(2-\lambda)\\
w^4(\tiny{\yng(1)};0) & = & -\frac{4}{5}  \,\sigma^2\, (1+\lambda) (2+\lambda) (3+\lambda) \qquad 
& w^4(0;\tiny{\yng(1)}) & = & - \frac{4}{5}\,  \sigma^2\, (1-\lambda)(2-\lambda) (3-\lambda) 
\end{array}
\end{equation}
and
\begin{eqnarray}
w^5(\tiny{\yng(1)};0) & = & -\tfrac{32}{35} \, i \, \sigma^3\, (1+\lambda) (2+\lambda) (3+\lambda) (4+\lambda)\ , \\
w^5(0;\tiny{\yng(1)}) & = &  \tfrac{32}{35}\, i\, \sigma^3 \, (1-\lambda)(2-\lambda) (3-\lambda) (4-\lambda) \ , \\[2pt]
w^6(\tiny{\yng(1)};0) & = & \tfrac{64}{63}\, \sigma^4\,(1+\lambda)(2+\lambda)(3+\lambda)(4+\lambda) (5+\lambda) \ , \\
w^6(0;\tiny{\yng(1)}) & = &  \tfrac{64}{63}\, \sigma^4\, (1-\lambda)(2-\lambda)(3-\lambda)(4-\lambda) (5-\lambda)  \ ,
\end{eqnarray}
from which we deduce
\be
w^2(\phi) =1 \ , \qquad
w^3(\phi) = 4\, i \sigma \lambda  \ , \qquad
w^4(\phi) = -\frac{48}{5} \sigma^2 (1+\lambda^2) \  ,
\ee
and 
\be
w^5(\phi) = -\frac{128}{7} \, i \, \sigma^3 \, \lambda (5+\lambda^2) \ , \quad
w^6(\phi) = \frac{640}{21}\, \sigma^4\, (8+15 \lambda^2 + \lambda^4) \ .
\ee
Together with the values for the structure constants (\ref{structure}),
it is then straightforward to check that eqs.~(\ref{w41}) -- (\ref{w61}) are indeed satisfied. This gives strong
support to the assertion that $\phi$ indeed satisfies all the requirements in (\ref{cond1}). We shall also be 
able to confirm this using different methods for the special cases of $\lambda=0$ and $\lambda=1$, see 
sections~\ref{sec:pertmu0} and \ref{sec:pertmu1} below.

\subsection{The analysis for the non-linear ${\cal W}_{\infty}[\lambda]$ case}

The above analysis was done in the 't~Hooft limit, but we may ask whether the situation for the  
${\cal W}_{\infty}[\lambda]$ algebra at finite $c$ would be different. We have repeated the analysis of 
(\ref{cond1}) for this case, using the explicit form of the quantum algebra ${\cal W}_{\infty}[\lambda]$ 
as given in \cite[Appendix~A]{Gaberdiel:2012ku} (see also \cite{MC}). While there are $\frac{1}{c}$ 
corrections, e.g.\ (\ref{w41}) and (\ref{w51}) become
\begin{eqnarray}
w^4 & = & \frac{3}{4} \Bigl( \, \frac{5\, (c-2)}{5c+22} \, N_3 + (w^3)^2 \Bigr) \\[2pt]
w^5 & = & \frac{1}{15} \Bigl( 10 \, w^4 w^3 + \frac{56 (c-6)}{(7c+114)} \, \frac{N_4}{N_3} \, w^3 \Bigr) \ ,
\end{eqnarray}
we have found that the general structure is largely unmodified, i.e.\ there continues to be a one-parameter family of such
perturbing fields (that are characterised by the $W^3_0$ eigenvalue $w^3$). In order to find
the analogue of $\phi$ in this context we have demanded in addition that the representation
generated from $\Phi$ has the same character as that of $\phi$, i.e.\ 
\be
\chi_{\phi} = \frac{q}{(1-q)^2} \, \prod_{s=2}^{\infty} \prod_{n=s}^{\infty} \frac{1}{(1-q^n)} 
= q^1 \Bigl(1 + 2 q + 4 q^2 + 7 q^3 + \cdots\Bigr) \ .
\ee
In particular, this means that there are only two linearly independent states at the first
descendant level, i.e.\  the representation possesses many null states, e.g.\ a null state of the form
$(W^4_{-1} + \alpha W^3_{-1} + \beta L_{-1}) \Phi = 0 $, etc. Then there are only six solutions for $w^3$,
namely the roots of the sextic equation
\begin{align}
&21000(5c + 22)^2(c-1)N_3^3 N_4 
-253125 (5c + 22) (c^2-4) N_3^5 \cr
& - 5625(259c^3 + 2170c^2 + 8180c + 9752)N_3^4 (w^3)^2 
-12544(c + 2)(5c + 22)^2 N_4^2\, (w^3)^2 \cr
& \quad +8400(5c + 22)(31c^2 + 141c + 266) N_3^2 N_4\, (w^3)^2 \cr
& -1125(5c + 22)(35c^2 -192c - 524) N_3^3\, (w^3)^4 
+840(5c - 17)(5c + 22)^2N_3 N_4\, (w^3)^4 \cr
& -225(5c + 22)(5c^2 + 32c + 44)N_3^2\, (w^3)^6 = 0 \ . \label{4.23}
\end{align}
In the large $c$ limit, plugging in the values for $N_3$ and $N_4$ from (\ref{structure}), we obtain the solutions
\be\label{4.24}
w^3 = \pm \, 4 \, i\, \sigma\, \lambda \quad \hbox{or} \quad
w^3 = \pm \, 4 \, i\, \sigma\, (\lambda + 2) \quad \hbox{or} \quad 
w^3 = \pm \, 4 \, i\, \sigma\, (\lambda - 2) \ .
\ee
Note that the first two solutions correspond precisely to $\phi$ and its conjugate $\phi^\ast$ (i.e.\ the corresponding
state in $(\overline{\tiny{\yng(1)}};\overline{\tiny{\yng(1)}})$). We don't know the interpretation for the other four solutions.

\medskip

It is instructive to compare these general results to what happens at the special points $\lambda=0$ and $\lambda=1$
where we have free field realisations of the algebra, see Section~\ref{sec:mu0} and \ref{sec:mu1}, respectively. 
In particular, for these cases we can construct the analogue of $\phi$ explicitly, and show that it preserves indeed
the full ${\cal W}_{\infty}[\lambda]$ algebra to first order. Let us first consider the case of $\lambda=0$.

\subsection{The perturbing field at $\lambda=0$}\label{sec:pertmu0}

In the free fermion theory there is only one ${\rm U}(1)$-primary field of conformal dimension $(1,1)$
in the untwisted sector, namely
\be\label{Phifdef}
\Phi = \left(\bar{\psi}^{*i}\, \psi^i\right)\,  \left( \bar{\psi}^j\, \psi^{*j} \right)  - \frac{1}{N} J \bar{J}  = 
\left(\bar{\psi}^{*i}\, \psi^i\right)\,  \left( \bar{\psi}^j\, \psi^{*j} \right)  - \frac{1}{N}  
\left({\psi}^{*i}\, \psi^i\right)\,  \left( \bar{\psi}^{\ast j}\, \bar{\psi}^j \right) \ , 
\ee
where $J$ and $\bar{J}$ are the holomorphic and anti-holomorphic ${\rm U}(1)$-currents, 
respectively. In terms of the continuous orbifold description it corresponds to the field
\be\label{Phico}
\Phi  = \sum_a J^a \bar{J}^a \ ,
\ee
as one confirms using the identity of the representation matrices in the fundamental representation
\be
\sum_a t^a_{ij} t^a_{kl} = \delta_{il}\, \delta_{jk} - \frac{1}{N}\delta_{ij} \delta_{kl} \ .
\ee
The perturbation by this field leaves the ${\cal W}_{\infty}[\lambda=0]$ currents holomorphic to first order. One way
to see this is to consider the perturbed action
\be
S = S_0 + g \int  d^2 z \, \Phi (z,\bar{z}) \ , 
\ee
where $S_0$ was defined in (\ref{feract}). This perturbation modifies the equations of motion of the free theory
(\ref{fereoms}) to 
\begin{equation}
\bar{\partial} \psi^i=g \,(\bar{\psi}^i K + \frac{1}{N}\, \psi^i \bar{J} )
\qquad  \qquad \bar{\partial} \psi^{*i} =-g \,( \bar{\psi}^{*i} \bar{K} + \frac{1}{N}\, \psi^{\ast i} \bar{J} ) \ ,
\end{equation}
where $K$ and $\bar{K}$ are defined as 
\be
K \equiv \bar\psi^{\ast j} \, \psi^j \ , \qquad \bar{K} \equiv \psi^{\ast j}\, \bar\psi^j \ . 
\ee
It is easy to check that the ${\rm U}(1)$ current remains conserved, $\bar{\partial} J = 0$, and 
the same is true for the stress energy tensor $T$ (and hence for $\tilde{T}$) since
\begin{eqnarray}
\bar{\partial}T & = & 2g\, 
\bigl[-(\bar{\psi}^{*}\cdot \partial \psi) \bar{K}-(\partial{\psi}^{*}\cdot \bar{\psi}) K+\partial(K\bar{K}) \bigr]  \nonumber \\
& = & 2g \, 
\bigl[ - g (\psi^{*} \cdot \psi) K \bar{K} + g (\psi^{*} \cdot \psi) \bar{K} K \bigr] = 0 \ ,
\end{eqnarray}
where the dot `$\cdot$' is a shorthand for the sum over $i$. On the other hand, one shows that
the higher spin currents $W^s$ (and hence $\tilde{W}^s$) are only preserved to first order in $g$,  
\begin{eqnarray}
\bar{\partial}W^{s}&=&-g\sum^{s-1}_{k=0}\sum^{s-1-k}_{p=1}\sum^{p}_{q=0}(-1)^k 
\binom{s-1}{k,p-q,q} \Bigl[\binom{s-1}{k}-(-1)^p\binom{s-1}{k+p}\Bigr]  \nonumber \\
& & \qquad  \times \ 
 (\partial^{s-1-k-p}\psi^{*}\cdot\partial^{p-q}\bar{\psi})(\partial^{q}\bar{\psi}^* \cdot \partial^{k}\psi)\ ,\label{classpert}
\end{eqnarray}
where 
\begin{equation}
{s-1 \choose k,p-q,q} = \frac{(s-1)!}{k! \, (p-q)! \, q!\, (s-1-k-p)! } \ .
\end{equation}
Since the sum over $p$ starts with $p=1$, we can apply the equations of motion at least once more, and 
find that $\bar\partial W^s = {\cal O}(g^2)$. 

Thus our perturbation by $\Phi$ should satisfy the conditions 
(\ref{cond1}) from above. In fact, using Wick's theorem repeatedly, one finds that 
\be
\tilde{W}_0 \Phi = 0 \ .
\ee
This solves indeed (\ref{4.23}) since, at $\lambda=0$, we have the relation
\begin{equation}
\frac{N_4}{N_3^2}  =   \frac{253125 (c^2-4)}{21000 (5c+22) (c-1)}  =
 \frac{675}{56} \, \frac{c^2-4}{(5 c+22)(c-1)} = \frac{75}{896} \left. \gamma^2\right|_{\lambda=0} \ ,
\end{equation}
where $\gamma^2$ was defined in (\ref{gammadef}) and (\ref{gammaex}). Furthermore, $\Phi$ 
corresponds to the solution with $w^3 = \pm 4\, i \, \sigma \left.\lambda\right|_{\lambda=0}=0$, i.e.\
it describes in the 't~Hooft limit precisely the left-right symmetric combination of the field $\phi$. 
\smallskip

While the perturbation by $\Phi$  in (\ref{Phifdef}) preserves the ${\cal W}_{\infty}[\lambda=0]$ currents to first order
in perturbation theory,
it does not do so to higher orders. One explicit way to see this is to use (\ref{classpert}) to determine 
\begin{equation}
\bar{\partial}\tilde{W}^{3}=6g^2\bigl[J\partial(\bar{K}K)-(\partial{J})\bar{K}K \bigr] \ ,
\end{equation}
which does not vanish.
We have also confirmed this conclusion by a direct perturbative
analysis. 

Another way to arrive at the same conclusion is to observe that it follows from 
the analysis of \cite{Chaudhuri:1988qb} that $\Phi$, given by (\ref{Phico}), 
 is {\em not} exactly marginal. Indeed, a current-current deformation is only exactly marginal
if all chiral currents that appear in the sum lie in an abelian subalgebra, and similarly for the
anti-chiral currents. However, this is clearly not the case for (\ref{Phico}).
Thus, in particular, the $T$ component of the stress energy tensor does not remain 
holomorphic to higher order in perturbation theory.
The same conclusion can also be reached by observing that in the free fermion description
we have the identification
\begin{equation}\label{ffid}
({\tiny\yng(1)};{\tiny\yng(1)}) \cong (0;0) \oplus ({\rm adj};0) \ ,
\end{equation}
as follows from \cite[eq.~(2.10)]{Gaberdiel:2011aa}. The perturbing field corresponds to the 
second representation, $({\rm adj};0)$, which is not exactly marginal.

\subsection{The perturbing field at $\lambda=1$}\label{sec:pertmu1}

For the free boson theory at $\lambda=1$, the only real singlet field (in the untwisted sector) that
has conformal dimension $(1,1)$ is 
\be
\Phi_1 = \partial \phi^{\, j} \, \bar\partial \bar\phi^{\, j}  +  \bar\partial \phi^{\, j} \, \partial \bar\phi^{\, j} \ .
\ee
It is proportional to the Lagrangian itself, and thus switching on this field only changes the `radius' of 
the bosons; in particular it therefore does not break the ${\cal W}_{\infty}[\lambda=1]$ symmetry. (It also
cannot deform $\lambda$ since otherwise ${\cal W}_{\infty}[\lambda]$ algebras with $\lambda\neq 1$ would also have a free 
boson realisation and hence must be linear, in contradiction with the results of 
\cite{Gaberdiel:2011wb}.\footnote{We thank Rajesh Gopakumar for this observation.})

Thus we should expect that $\Phi_1$ is again a solution of (\ref{4.23}). Using Wick's theorem, we have 
determined the $W^3_0$ eigenvalue of $\Phi_1$; in fact, the two terms in $\Phi_1$ (that are complex conjugates of one
another) have opposite $W^3_0$ eigenvalue, and thus $\Phi_1$ is only an eigenvector of $(W^3_0)^2$ with
eigenvalue
\be\label{wbos}
(w^3)^2 = 16 \ .
\ee
Together with (\ref{3.16}) one can easily check that (\ref{4.23}) is indeed satisfied for all values of $c$, i.e.\ 
all values of $k$. We should also mention that comparing (\ref{3.16}) to (\ref{structure}) it follows 
that $\sigma^2=-1$ in the conventions of Section~\ref{sec:mu1}. Thus (\ref{wbos}) corresponds again
to the first eigenvalue in (\ref{4.24}) at $\lambda=1$, i.e.\ $\Phi_1$ can be identified with the left-right symmetric
version of $\phi + \phi^\ast$ in the 't~Hooft limit.

The fact that the perturbing field is trivial can also be understood from the point of view of the minimal models. 
At $\lambda=1$, all eigenvalues of the fields of the form $(0;\Lambda_-)$ vanish, and hence the analogue of 
\cite[eq.~(2.10)]{Gaberdiel:2011aa} is  
\begin{equation}
(\Lambda_1;\Lambda_2) \cong (\Lambda_1;0) \ .
\end{equation}
The analogue of (\ref{ffid}) is therefore 
\begin{equation}
({\tiny\yng(1)};{\tiny\yng(1)}) \cong ({\tiny\yng(1)};0) \ , 
\end{equation}
i.e.\ the perturbing field agrees indeed with the `scalar' field $\partial \phi^j \bar\partial \bar\phi^j$.

\section{The effect of the perturbation}\label{sec:pert}

As we have seen in Section~\ref{sec:perttH}, in the 't~Hooft limit 
the $\phi$ `descendant' of the $(\tiny\yng(1);\tiny\yng(1))$  representation (which decouples
from the ground state in the 't~Hooft limit)
defines a perturbation that leaves the full set of ${\cal W}_{\infty}[\lambda]$ currents holomorphic 
at first order in perturbation theory. Thus we can ask how the perturbation by the corresponding left-right symmetric field 
$\Phi$ (i.e.\  the combination of $\phi$ with its right-moving analogue) changes the underlying ${\cal W}_{\infty}[\lambda]$ 
theory. 

Since the perturbation only has the desired properties in the 't~Hooft limit, we need to be  careful about how
precisely the limit is defined. Let us denote by 
$\omega$ the ground state of the $(\tiny\yng(1);\tiny\yng(1))$  representation in the ${\cal W}_{N,k}$ minimal model, 
and by $\omega^{\ast}$ its conjugate, i.e.\ the 
ground state of the $(\overline{\tiny\yng(1)};\overline{\tiny\yng(1)})$ representation. At finite $(N,k)$ we define,
following \cite{Chang:2013izp} 
\begin{equation}\label{limit}
\Phi = \frac{1}{2h_\omega}\,   \bar{L}_{-1}L_{-1} \omega \ ,\qquad \qquad  
\omega =  \frac{1}{2h_\omega}\,  \bar{L}_1 L_1 \Phi \ , 
\end{equation}
and then take the $N\rightarrow \infty$ limit, keeping as before 
$\lambda = \frac{N}{N+k}$ fixed, see eq.~(\ref{lambdadef}).\footnote{Another choice 
for the limit theory is to demand that as $N\rightarrow \infty$,  $\omega$ becomes null whereas $\Phi$ stays in the spectrum and has a unit norm. This requires $\Phi = \frac{1}{(2h_{\omega})^2} \bar{L}_{-1}L_{-1} \omega$ and $  
\omega = \bar{L}_1 L_1 \Phi $ \cite{Gaberdiel:2011zw}. }
Then both
$\omega$ and $\Phi$ have unit norm in the limit, but become disconnected in the sense that 
\begin{equation}
\bar{L}_{-1}L_{-1} \omega = 0 \ , \qquad \bar{L}_1 L_1 \Phi = 0 \ .
\end{equation}
In particular, $\Phi$ is therefore a primary non-descendant field, and it makes sense to perturb with it. 
Actually, since we should  perturb with a real field we shall consider the perturbation by $P=\Phi + \Phi^*$, 
where 
\be\label{Phidef}
\Phi= \lim_{N\rightarrow \infty \atop \lambda \textrm{ fixed}} \,
\frac{1}{2h_\omega}\,\bar{L}_{-1}L_{-1} \omega \ , \qquad 
\Phi^*= \lim_{N\rightarrow \infty  \atop \lambda \textrm{ fixed}} \, \frac{1}{2h_\omega}\,\bar{L}_{-1}L_{-1} \omega^* \ . 
\ee
Since the four-point function of $P$ with itself does not factorise over the identity channel, the perturbation
by $P$ will not be exactly marginal \cite{Cardy:1987vr}. 
However, as in \cite{Chang:2013izp} there are suitable holomorphic
descendants of $\omega$ (and its higher powers) that begin to mix with the stress energy tensor in perturbation
theory, and it is plausible that a suitable linear combination of them will remain holomorphic. In any case,
we shall only work to first order in perturbation theory, where the theory remains conformal.

\subsection{The structure of the perturbation theory}\label{sec:structure}

In order to study the effect of the perturbation by $P$, one could try to study the behaviour of the
structure constants of the ${\cal W}_{\infty}[\lambda]$ algebra (i.e.\ the correlators of the holomorpic currents)
under the perturbation by $P$. However,
this is quite delicate since the perturbation directly affects the holomorphic correlators only at second
order in perturbation theory: since $P$ has conformal dimension $(1,1)$, the `right-moving'
conformal dimension must be soaked up by at least one other field with non-zero  right-moving
conformal dimension. In fact, there is a quite generic argument that shows that $\lambda$ can never be changed by
an exactly marginal operator $P$.\footnote{We thank Stefan
Fredenhagen for explaining this to us.} In order to see this, suppose that we have a family of ${\cal W}$-algebras
(parametrised by $\lambda$), and suppose that we could change $\lambda$ by switching on a 
perturbation by an exactly marginal field $P$ 
with coupling constant $g$. As we have just explained, the first order perturbation must always vanish,
 thus the derivative of all ${\cal W}$-algebra correlators with respect to $g$ 
vanishes when evaluated at $g=0$ --- and this holds for all values of $\lambda$.  
But if the perturbation by $g\, P$  just changes $\lambda$, then the fact that this 
holds for all values of $\lambda$ implies that it also holds for all values of $g$, i.e.\ 
that the derivative of the ${\cal W}$-algebra correlators with respect to  $g$ vanishes 
for all values of $g$. But then this means that these correlators are actually independent
of $g$, i.e.\ that the ${\cal W}$-algebra does not change under the perturbation.

In our case the situation is  different since $P$ is not exactly marginal, and hence the previous
argument does not apply. However, it highlights the difficulty in trying to determine the change of $\lambda$
directly from the correlators  of the holomorphic currents. We shall therefore follow a different route: we will compute 
 the conformal dimension of the simplest representation ${\cal O}\equiv (\tiny\yng(1);0)$  (which has conformal dimension $h=\frac{1}{2} (1+\lambda)$ at $\mathcal{O}(g^0)$) in the perturbed theory, and read off the effect of the perturbation from the change of $h$. The computation is an analysis of operator-mixing following \cite{Zamolodchikov:1987zf} (see also \cite{Zamrev}) which we shall now outline. 
\smallskip
 
Let us consider all the fields $\{\mathcal{O}_i, \mathcal{O}^*_i\}$  with conformal dimension $h=\frac{1+\lambda}{2}$ in the unperturbed theory, where ${\cal O}_i$ and ${\cal O}^*_i$ are a conjugate pair. Define $\mathcal{M}$ as the matrix of their two-point functions. Because these two-point functions always couple 
conjugate pairs together, the structure of $\mathcal{M}$ is of the form 
\be
{\cal M} = \textbf{M} \otimes \left( \begin{array}{cc} 0 & 1 \cr 1 & 0 \end{array} \right)  \ .
\ee
As we shall see, we only need to know the eigenvalues of $\mathcal{M}$; therefore it is enough to concentrate on the matrix $\textbf{M}$, i.e.\ we shall simply write
${\cal O}_i$ for the pair $\{{\cal O}_i,{\cal O}^*_i\}$,
\begin{equation}
\textbf{M}_{ij}=\langle \mathcal{O}_{i}(z_1,\bar{z}_1) \mathcal{O}_j(z_2,\bar{z}_2)\rangle \ .
\end{equation}
At $\mathcal{O}(g^0)$, the matrix $\textbf{M}$ is diagonal
\begin{equation}
\textbf{M}^{(0)}_{ij}=\langle \mathcal{O}_{i}(z_1,\bar{z}_1) \mathcal{O}_j(z_2,\bar{z}_2)\rangle^{(0)}=\frac{1}{|z_1-z_2|^{4 h}}
\, \delta_{ij} 
\quad \textrm{and }\ln \textbf{M}^{(0)}=(-4h)\ln |z_1-z_2|\, \mathbf{1} \ .
\end{equation}
At order $g$, two representations $(\mathcal{O}_{i}$ and $\mathcal{O}_{j})$ that are related to each other by adding 
or subtracting a box can start to have a non-zero two-point function
\begin{eqnarray}
\langle \mathcal{O}_i(z_1,\bar{z}_1) \mathcal{O}_j(z_2,\bar{z}_2)\rangle^{(1)}& = & 
g\int d^2w \, \langle \mathcal{O}_i(z_1,\bar{z}_1) \mathcal{O}_j(z_2,\bar{z}_2)\, P(w,\bar{w})\rangle \nonumber \\
& = & g \,  \frac{1}{|z_1 -z_2|^{4h}}\ln{|z_1-z_2|}\, \mathbf{P}_{ij} \ .
\label{mixing}
\end{eqnarray}
Thus, to $\mathcal{O}(g)$, the mixing matrix $\mathbf{M}$ is
\begin{equation}\label{mixingmatrix}
\mathbf{M}=\mathbf{M}^{(0)}+g \,\mathbf{P}+\mathcal{O}(g^2)=\frac{1}{|z_1-z_2|^{4 h}}
\Bigl(\mathbf{1}+g\, \mathbf{P}\ln{|z_1-z_2|}+\mathcal{O}(g^2) \Bigr) 
\end{equation}
with $\ln \mathbf{M}=[ (-4h)\mathbf{1}+g\mathbf{P} \, ]\ln |z_1-z_2|+\mathcal{O}(g^2)$. 
In particular, after the 
perturbation is turned on, the two-point function matrix is no longer diagonal, i.e.\ 
the fields $\{\mathcal{O}_i, \mathcal{O}^*_i\}$ we started with are no longer conformal eigenstates. 
The new conformal eigenstates in the perturbed theory at $\mathcal{O}(g)$ are the eigenvectors of the 
mixing matrix defined as 
\begin{equation}\label{mixmatg}
 M=(-4h)\, \mathbf{1}+g\, \mathbf{P} \ ,
\end{equation}
and their conformal dimensions are the corresponding eigenvalues divided by $(-4)$.

Now in order to study the effect of the perturbing field $P$ on $(\tiny\yng(1);0)$, we first need to identify all states that 
are `mixed' together  at order $g$; a necessary condition for this is that they have 
conformal dimension 
$h=\frac{1}{2} (1+\lambda)$ in the 't~Hooft limit. To enumerate these fields, first recall that the conformal dimension of 
$(\Lambda_{+};\Lambda_{-})$ at finite $(N,k)$ can be written in terms of the quadratic Casimir $C_2$ of $\mathfrak{su}(N)$  as 
\begin{equation}\label{hfiniteNk}
h(\Lambda_{+};\Lambda_{-})=\frac{C_2(\Lambda_{+})}{N+k}+\frac{C_2(\Lambda_{0})}{N+1}-\frac{C_2(\Lambda_{-})}{N+k+1}+n \ ,
\end{equation}
where $\{\Lambda_{+}, \Lambda_{0}, \Lambda_{-}\}$ are highest weight representations of 
$\{ \mathfrak{su(}N)_{k}, \mathfrak{su}(N)_{1}, \mathfrak{su}(N)_{k+1}\}$, respectively. They satisfy the constraint that, as
a weight of $\mathfrak{su}(N)$, $\Lambda_{+}+\Lambda_{0} -\Lambda_{-}$ lies in the root lattice of $\mathfrak{su}(N)$. 
Furthermore, $n$ is the `height' at which $\Lambda_{-}$ appears in $\Lambda_{+}\otimes \Lambda_{0}$. The quadratic
Casimir has the large $N$ expansion (that is exact even at finite $N$)
\begin{equation}\label{casimir2}
C_2(\Lambda)=N\, \frac{B(\Lambda)}{2}+\frac{D(\Lambda)}{2}- \frac{B(\Lambda)^2}{2 N}  \ ,
\end{equation}
where $B(\Lambda)$ is the number of boxes of $\Lambda$,
$B(\Lambda) = \sum_i r_i = \sum_j  c_j$,
and $D(\Lambda)$ is defined as 
$D(\Lambda) = \sum_i r_i^2-\sum_j c_j^2$,  with $r_i$ and $c_j$ being the number of boxes in the $i$'th row and $j$'th 
column, respectively. 
Expressed in terms of $B$ and $D$, the large $N$ expansion of the conformal dimension (\ref{hfiniteNk}) is then
\begin{equation}\label{hexpand}
h(\Lambda_{+};\Lambda_{-})=\frac{B_{0}+(B_{+}-B_{-})\lambda}{2}+n
+\frac{1}{2N}\Bigl [(D_{0}-B_{0}) +(D_{+}-D_{-})\lambda+B_{-}\lambda^2\Bigr]
+\mathcal{O}\left(\frac{1}{N^2}\right) \ ,
\end{equation}
where $B_{+}$ is a shorthand for $B(\Lambda_{+})$, and similarly for the others.\footnote{Actually,
in the large $N$ limit, $B(\Lambda)$ is in general the sum of the number of boxes and anti-boxes
of $\Lambda$, see e.g.\ \cite{Gaberdiel:2011zw}. In particular, $B_0=1$ in (\ref{5.13}) corresponds
to $\Lambda_0$ being the anti-fundamental representation with one anti-box.} Therefore for the state 
$(\Lambda_{+};\Lambda_{-})$ to have conformal dimension $\frac{1+\lambda}{2}$ in the 't Hooft limit we need
\begin{equation}\label{5.13}
B_{0}=1, \qquad  B_{+}-B_{-}=1, \qquad \textrm{and } \quad n=0 \ ,
\end{equation}
which means that 
$\Lambda_+ \subset \Lambda_- \otimes \tiny\yng(1)$. For these representations --- in the following
we shall refer to them sometimes as the `scalar-like' representations --- the conformal
dimension then equals
\be\label{generalw}
h(\Lambda_+;\Lambda_-) = \frac{1}{2} (1+\lambda) + \frac{\lambda^2 B_{-} - 1 }{2N} 
+ \frac{D_+- D_-}{2N} \, \lambda 
+ {\cal O}\Bigl(\frac{1}{N^2}\Bigr) \ .
\ee
Note that this formula is only correct for $\lambda<1$; for $\lambda=1$, we need to take
$N\rightarrow \infty$ at finite $k$, and then an expansion in inverse powers of $N$ does not make sense. 
Instead, we should then consider an expansion in inverse powers of $k$, i.e.\ we should
replace 
\begin{equation}\label{Nkrep}
\frac{1}{N} \ \mapsto  \ \frac{(1-\lambda)}{\lambda}\,  \frac{1}{k} \ .
\end{equation}
\smallskip

\begin{figure}[ht]
\centering
\includegraphics[width=.85\textwidth]{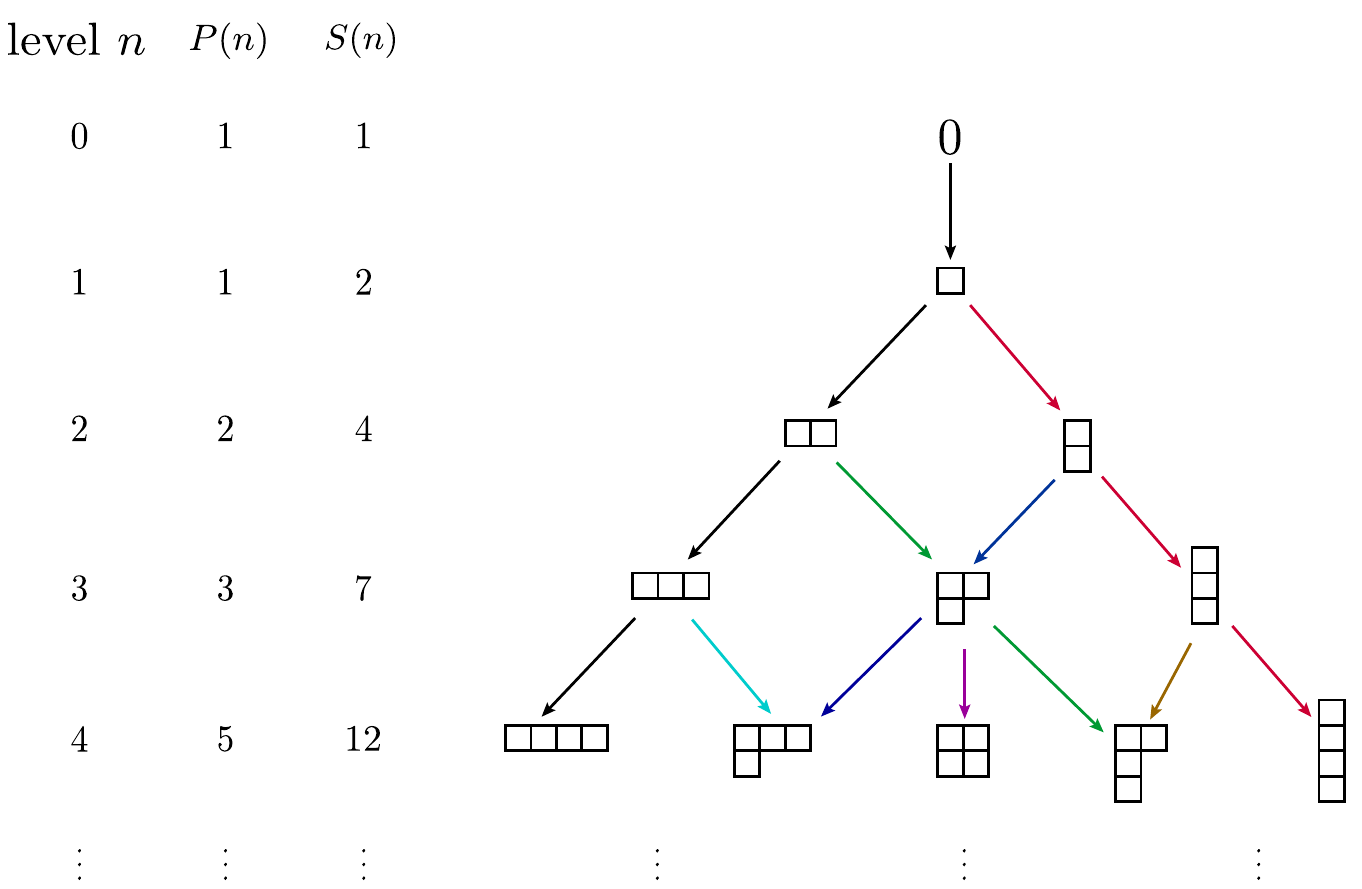}
\caption{Young lattice.}
\label{Young}
\end{figure}

The property that $\Lambda_+ \subset \Lambda_- \otimes \tiny\yng(1)$ implies that the fields of interest are 
in one-to-one correspondence with the edges of the  Young lattice (see e.g.\ \cite{combinatorics} for an introduction
to some of its basic properties --- the different colours of the arrows in Fig. \ref{Young} will be explained below).

Recall that the number of Young tableaux at level $n$ (i.e.\ with $n$ boxes) equals $P(n)$ (the partition number of $n$), while
the number of edges at level $n$ (i.e.\ edges from level $n$ to level $n+1$) is $S(n)$ defined by  
\be\label{SP}
S(n) = \sum_{k=0}^{n} P(k) \ .
\ee
Their generating functions are 
\begin{equation}
\sum^{\infty}_{n=0} P(n)\, x^n=\prod^{\infty}_{m=1}\frac{1}{1-x^m}
\qquad\qquad 
\sum^{\infty}_{n=0} S(n)\, x^n=\frac{1}{1-x} \prod^{\infty}_{m=1}\frac{1}{1-x^m} \ . 
\end{equation}
 $S(n)$ counts the number of scalar-like fields for which $\Lambda_-$ has $n$ boxes, and it grows exponentially 
 with $n$.\footnote{Indeed, we have 
 $P(n)\sim \frac{1}{4 \sqrt{3}n}\exp\bigl({\pi\sqrt{\frac{2 n}{3}}}\bigr)$ as $n \to \infty$.} Therefore  there are 
 infinitely many such fields in the 't~Hooft limit, and we 
need to deal with an infinitely-degenerate operator mixing problem. However, as we shall see below in 
Section~\ref{sec:eigenv}, the perturbation analysis has a lot of 
structure, and we can therefore understand at least its qualitative features in some detail.

\subsection{The mixing matrix}

Next we want to determine the mixing matrix $M$ explicitly. Its off-diagonal entries 
can be computed from the integral
\begin{equation}
\int d^2w \, \langle {\cal O}_i(z_1,\bar{z}_1) {\cal O}_j(z_2,\bar{z}_2)P(w,\bar{w})\rangle
=\mathbf{P}_{ij}  \frac{1}{|z_1-z_2|^{4h}}\ln{|z_1-z_2|} \ ,
\end{equation}
where $h= \frac{1}{2}(1+\lambda)$ is the conformal dimension at $\mathcal{O}(g^0)$.
As explained above, the correlators in the integrand are first to be evaluated at finite $N$ and $k$, 
with $P=\Phi+\Phi^*$ as given in (\ref{Phidef}); the large $N$, $k$ limit is then only taken at the end.
At finite $N$ and $k$ and up to ${\cal O}(\frac{1}{N})$, the three-point function of ${\cal O}_{1}$ and
${\cal O}_{2}$ with $\omega$ equals
\begin{equation}\label{3ptomega}
\langle {\cal O}_{1}(z_1,\bar{z}_1) {\cal O}_2(z_2,\bar{z}_2)\, \omega(w,\bar{w}) \rangle
=C_{{\cal O}_1{\cal O}_2\, \omega}
\left| \frac{1}{(z_1-z_2)^{2h+\frac{1}{N}\Delta_1}(z_2-w)^{\frac{1}{N}\Delta_2}(z_1-w)^{\frac{1}{N}\Delta_3}}\right|^2 \ ,
\end{equation}
where $\Delta_{i}=\delta_i+\delta_{i+1}-\delta_{i+2}$, and $\delta_i$ (with $i\equiv i+3$) are the $\mathcal{O}(\frac{1}{N})$ 
corrections to the conformal dimensions
\begin{eqnarray}
{\cal O}_1: \ \ h_1=h+\frac{\delta_1}{N} \quad \qquad {\cal O}_2: \ \ h_2=h+\frac{\delta_2}{N} \quad \qquad
\omega: \ \ h_3=\frac{\delta_3}{N}=\frac{\lambda^2}{2N}  \ . 
\end{eqnarray}
In order to deduce from this the correlator corresponding to $\Phi$, we then apply  
$\frac{1}{2h_{\omega}}\bar{L}_{-1}L_{-1}$ to (\ref{3ptomega}) and obtain
\begin{equation}
\langle {\cal O}_1(z_1,\bar{z}_1) {\cal O}_2(z_2,\bar{z_2})\, \Phi(w,\bar{w}) \rangle\\
=\frac{4}{N\lambda^2}C_{{\cal O}_1 {\cal O}_2 \, \omega}
\left|\frac{\frac{\Delta_2}{z_2-w}+\frac{\Delta_3}{z_1-w}}{(z_1-z_2)^{2h+\frac{1}{N}
\Delta_1}(z_2-w)^{\frac{1}{N}\Delta_2}(z_1-w)^{\frac{1}{N}\Delta_3}}\right|^2 \ .
\end{equation}
Finally, we take the large $N$ limit and evaluate the $w$-integral to obtain
\begin{equation}\label{integralreg}
\int d^2 w\, \langle {\cal O}_1(z_1,\bar{z}_1) {\cal O}_2(z_2,\bar{z}_2)\, \Phi(w,\bar{w})\rangle
=\Bigl(-\frac{16 \pi}{N}\cdot \frac{\Delta_2 \Delta_3}{\lambda^2}\cdot C_{{\cal O}_1 {\cal O}_2 \, \omega}\Bigr)
\frac{1}{|z_1-z_2|^{4h}}\ln|z_1-z_2| \ , 
\end{equation}
where $\Delta_2 \Delta_3=(\frac{\lambda^2}{2})^2-(\delta_1-\delta_2)^2$; here we have used the 
identity
\begin{equation}
\int d^2 w \frac{1}{(w-z_1)(\bar{w}-\bar{z}_2)} = 2 \pi \ln r\, \Bigr\vert^{\infty}_{| z_1 - z_2 |} \ .
\end{equation} 
and dropped the $\ln \infty$ term.\footnote{The singularity $\ln \infty$ is due to taking the $N \to \infty$ limit 
before doing the integral; the integral (\ref{integralreg}) does not suffer from an IR divergence 
at finite $N$.}

Thus the $\mathbf{P}_{ij}$ entry in the mixing matrix between ${\cal O}_i$ and ${\cal O}_j$ is 
\begin{equation}\label{Pij}
\mathbf{P}_{ij}=\frac{16 \pi}{N}  \cdot C_{{\cal O}_i{\cal O}_j\, \omega} \cdot 
\Bigl(\frac{(\delta_i-\delta_j)^2}{\lambda^2}  - \frac{\lambda^2}{4}\Bigr) \ .
\end{equation}
The coefficients $\delta_i$ and $\delta_j$ are read off from (\ref{generalw}), and for the calculation of the structure constants
$C_{{\cal O}_i{\cal O}_j\, \omega}$ we can use the results of \cite{Chang:2011vka,Jevicki:2013kma},
see also \cite{Papadodimas:2011pf} for earlier work. In particular,
some of them were calculated explicitly in \cite{Chang:2011vka} using the Coulomb gas 
approach. Based on the large $N$ factorisation properties of these structure constants, \cite{Jevicki:2013kma} wrote down 
an effective Hamiltonian that captures the exact spectrum and cubic interactions in the 't~Hooft limit. Since we only need 
the large $N$ result, we can therefore directly use this effective Hamiltonian point of view.

We have worked out the explicit coefficients for the low-lying representations up to level~$4$; some of the details are spelled
out in Appendix~\ref{sec:mixmat}.

\subsection{The eigenvalue problem}\label{sec:eigenv}

As we have reviewed above, the
eigenvalues of the mixing matrix $M$ (\ref{mixmatg}) give the perturbed  conformal
dimensions of the states under consideration. Thus we need to  study the eigenproblem
\begin{equation}\label{eigen1}
M\cdot \vec{F}=\rho\, \vec{F} \ .
\end{equation} 
In the strict $N\rightarrow \infty$ limit $\vec{F}$ is an infinite-dimensional vector, 
and hence we expect $\rho$ to take a continuum of eigenvalues. Nevertheless, as we shall now explain,
the structure of the eigenvalues can be identified naturally with that of the conformal dimensions of the scalar-like states in
the 't~Hooft limit.
\smallskip

\noindent To start with we observe that the matrix $M$ has the form of a block Jacobi matrix
\begin{equation}\label{Jacobi}
M(\{A_n\}, \{B_n\}) = 
\begin{pmatrix}
A_1 & B_1 & 0 & 0 & 0 & \ldots \\
B^{T}_1 & A_2 & B_2 & 0 & 0 & \ldots \\
0 & B^T_2 & A_3 & B_3 & 0 & \ldots \\
0 & 0 & \ddots & \ddots & \ddots & 0
\end{pmatrix}
\end{equation}
with
\begin{equation}
A_n=-4h \, \mathbf{1}_{S(n-1)\times S(n-1)} \ , \qquad B_{n}= S(n-1)\times S(n) \, \textrm{matrix of rank $S(n-1)$ .}
\end{equation}
The first few explicit expressions for $B_n$ are given in 
Appendix~\ref{sec:mixmat}. At low levels (for which we have worked them out, i.e.\ up to level $4$), the matrix 
$B_n$ has rank $S(n-1)$ provided that $\lambda\neq 0$. This is also what one should expect generically, and we 
therefore conjecture that this will continue to be true for all $n$. 
 
Under this assumption the eigenvector problem (\ref{eigen1}) can be 
solved recursively, generalising the method of finding the eigenvalues of 
a scalar Jacobi matrix (which we review in Appendix~\ref{sec:mixmatscalar}).
We first decompose the eigenvector $\vec{F}$ into 
$\vec{F}=\{\vec{f}_1,\vec{f}_2,\ldots\}$, with $\vec{f}_n$ being a vector of dimension $S(n-1)$. Then for any given 
eigenvalue $\rho$, we have to solve the recursive relations
\begin{equation}\label{genrec}
B_n \cdot \vec{f}_{n+1}=(\rho+4h) \vec{f}_{n}-B^T_{n-1}\cdot \vec{f}_{n-1} \ ,  \qquad n\in \mathbb{N}_{\geq 1} \ ,
\end{equation}
where we have set $\vec{f}_0\equiv 0$. 
Unlike the scalar Jacobi case for which the equation, at each level, is a scalar equation, now the equation at 
level $n$ is a matrix equation with $S(n-1)$ components for the $S(n)$ unknowns in $\vec{f}_{n+1}$. 
However, since  $B_n$ has rank $S(n-1)$, we can always find {\em a} solution
for $\vec{f}_{n+1}$, and hence an eigenvector with eigenvalue $\rho$. This solution is, however, not unique --- in 
fact the kernel of $B_n$ has dimension
$S(n)-S(n-1)=P(n)$, and thus, at every level $n$, we obtain $P(n)$ new families of solutions with
$\vec{f}_{n+1}\neq 0$ but $\vec{f}_i=0$ for $i\leq n$. 

In order to illustrate the structure of these various solutions let us describe some simple examples. Consider first the 
eigenvector that involves the original scalar representation $(\tiny\yng(1);0)$,
i.e.\ the eigenvector with $f_1=1$. In this case, (\ref{genrec}) is to be solved with the initial condition
\begin{eqnarray}\label{blockrecursion0}
f_1=1\ . 
\end{eqnarray}
We may choose to supplement 
the recursion relation (\ref{genrec}) with the requirement that $\vec{f}_{n+1}$ is 
orthogonal to the 
kernel  of $B_n$,
\begin{equation}\label{ker}
\vec{f}_{n+1}^{\ T}\cdot g_n=0 \qquad \forall g_n\in {\rm ker}(B_n) 
\end{equation}
in order to guarantee that at level $n$ the corresponding eigenvector is orthogonal to the new families of 
eigenvectors that will appear at that level.
These $P(n)$ equations, together with the $S(n-1)$ equations from (\ref{genrec}), then uniquely determine the 
$S(n)$ components of $\vec{f}_{n+1}$ for all $n\geq 1$.
This construction works for arbitrary $\rho$, and thus we conclude that the eigenvalue spectrum is continuous. This is 
consistent with the fact that, in the strict $N\rightarrow \infty$ limit, 
there are infinitely many states that are `mixed' with $(\tiny\yng(1);0)$ via $P$. 
\smallskip

In constructing the solution with $f_1=1$ we have in effect identified from the 
infinite mixing matrix $M$ (\ref{mixmatg})  a sub-matrix (which is also infinite) that describes the mixing of 
$(\tiny\yng(1);0)$ with all states that couple to it at $\mathcal{O}(g)$.  
For example, the condition (\ref{ker}) with $n=1$ identifies a specific linear combination 
$\vec{f}^{\; (1)}_2$ of the two states $(\tiny\yng(2);\tiny\yng(1))$ and $(\tiny\yng(1,1);\tiny\yng(1))$ at level $2$, 
namely the linear combination that mixes with $(\tiny\yng(1);0)$ at order $g$, and similarly at the 
higher levels. However, since there are two scalar-like states at level $2$, 
we may construct a second state $\vec{f}^{\; (2)}_2$ from $(\tiny\yng(2);\tiny\yng(1))$ and $(\tiny\yng(1,1);\tiny\yng(1))$
that is orthogonal to $\vec{f}^{\; (1)}_2$ and hence lies in the kernel of $B_1$. It will generate a new 
family of solutions; this is to say, we can repeat  the above procedure replacing $(\tiny\yng(1);0)$ by $\vec{f}^{\; (2)}_2$,
i.e.\  by imposing the initial conditions
\begin{eqnarray}\label{blockrecursion}
f_1=0\ ,\qquad \vec{f}_2=\vec{f}^{\; (2)}_2\ .
\end{eqnarray}
The eigenvectors of this second sub-matrix then involve the states that mix with the 
linear combination of $(\tiny\yng(2);\tiny\yng(1))$ and $(\tiny\yng(1,1);\tiny\yng(1))$ given by $\vec{f}^{\; (2)}_2$
at order $g$. Again, we can find the solution for the $\vec{f}_n$ recursively, requiring as before (\ref{ker}) for 
$n\geq 2$, and since this can be done for any 
choice of $\rho$, it will also lead to a continuum of eigenvalues.

It should now be clear how to proceed in general: at level $n$, we find $P(n)=S(n)-S(n-1)$ new families of solutions 
for which $\vec{f}_{n+1}\neq 0$ lies in the kernel of $B_n$ and $\vec{f}_j=0$ for $j\leq n$. For each choice of 
$\vec{f}_{n+1}$ we can then construct an eigenvector
recursively for any value of $\rho$. Thus each of these families will also give rise to a continuum of eigenvalues in the
't~Hooft limit. Altogether we therefore get at each level $n$, $P(n)$ families of eigenvectors, where each family
has in turn a continuum of eigenvalues.
\smallskip

So far we have studied the eigenvalue problem in the strict $N\rightarrow \infty$ limit. It is natural to ask whether it is possible to 
formulate (and hopefully answer) the same question at large but finite $N$. At finite $(N,k)$, the number of scalar-like states is 
finite ($\sim \min(N,k)$) and correspondingly the spectrum is no longer a continuum. In fact, as is explained in 
Appendix~\ref{sec:mixmatscalar}, it is straightforward to determine the eigenvalues of the finite problem by similar techniques. 
The resulting eigenvalues are then distributed symmetrically with respect to $h=\frac{1+\lambda}{2}$.

\subsection{The CFT interpretation}

As we have explained above, see eq.~(\ref{mixmatg}), 
each eigenstate with eigenvalue $\rho$ corresponds to a state in the spectrum with conformal dimension
\be\label{hrho}
h_\rho =  - \frac{\rho}{4} = h -  \frac{1}{4} (\rho  +4h)  \ , \qquad \hbox{i.e.} \quad \delta h = - \frac{1}{4} (\rho + 4h) \ . 
\ee
We now want to match $\delta h$ as computed from the perturbative analysis, 
to the change in conformal dimension of the primary operators of the
${\cal W}_{N,k}$ models in the 't~Hooft limit as we vary $(N,k)$. 
Recall that the 't~Hooft parameter and the central charge of the ${\cal W}_{N,k}$ models are given by 
$\lambda$ and $c_{N,k}$ in (\ref{lambdadef}) and (\ref{cNk}), respectively. As we modify 
$N\mapsto N+\delta N$ and $k\mapsto k+\delta k$, they change to first order (in the 't~Hooft limit) as
\be\label{dlam}
\delta \lambda = - \frac{1}{N}\lambda(\lambda-1) \delta N - \frac{1}{N} \lambda^2 \delta k 
\ee
and
\be\label{dc}
\delta c = (2\lambda+1)(\lambda -1)^2 \delta N + 2 \lambda^3 \delta k \ .
\ee
Furthermore, the $\gamma^2$ parameter, defined  in (\ref{gammaex}), can be expressed in terms
of $N$ and $k$ as 
\begin{equation}
\gamma^2  =  \frac{64 (k+1) (N-3) (N+1) (k+2 N) (3 k+2 N) (3 k+4 N+3)}{(N-2) (2 k+N) (2 k+3 N+2) 
\bigl( k (5 N+17)(k + 2 N+1) +22 N (N+1) \bigr)} \ ,
\end{equation}
and thus it changes as 
\be\label{dgam}
\delta \gamma^2 = -\frac{128 \lambda^2 (\lambda-1)}{N (\lambda-2)^2 (\lambda+2)^2} \, \delta N - 
\frac{128 \lambda^3}{N (\lambda-2)^2 (\lambda+2)^2} \, \delta k \ .
\ee
Given our general argument above (see the beginning of Section~\ref{sec:structure}) we should
expect that to first order  $\delta \gamma^2 =0$. Thus we conclude that $\delta k$ and $\delta N$ should be related as 
\be\label{dkN}
\delta k = -  \frac{(\lambda -1)}{\lambda} \, \delta N \ .
\ee
Note that then we also have $\delta \lambda=0$, and 
\be
\delta c = (1-\lambda^2)\,  \delta N \ .
\ee
Under this deformation the conformal dimensions of the scalar-like representations in eq.~(\ref{generalw}) 
change as 
\be\label{hdef}
\delta h (\Lambda_+;\Lambda_-) =-  \frac{1}{2}\, \frac{\delta N}{N}\, 
\Bigl[ \frac{\lambda^2 B_{-}-1}{N}   + \frac{(D_+- D_-)}{N}\, \lambda  \Bigr] \ .
\ee
This should now be compared to (\ref{hrho}) above. 

In order to do so, let us first explain how the structure of the answer is the same on both sides.
As we have seen above, the eigenstates of the mixing matrix $M$ organise themselves into families, where 
at each level $n$, $P(n)$ new families of eigenvectors of $M$ emerge. Each such family gives rise to a continuum
of eigenvalues $\rho$. 

From the viewpoint of the minimal model representations on the other hand, let us consider the 
`branches' of the Young lattice. Recall that the edges of the Young lattice are in one-to-one
corresponding to the scalar-like states $(\Lambda_+;\Lambda_-)$. A branch is a collection of edges,
starting from a given level and including one edge at each subsequent level. Given the structure of the Young 
lattice, see in particular eq.~(\ref{SP}), 
 it is clear that at each level $n$,  there are $P(n)$ edges that are not on any branch that emerges
before level $n$ and are the roots of $P(n)$ new branches. Thus the branches of the Young lattice are
in natural one-to-one correspondence to the families of eigenstates of $M$. 

One definite (and natural) way to construct these branches is as follows. Let us begin with the 
original scalar representation $(\tiny\yng(1);0)$,
and define its branch by picking one representation at each level (i.e.\ for each value of $B_-$),
namely the one  for which $\Delta D \equiv D_+ - D_-$ is maximal. (This is the branch described by
the black edges in Fig. \ref{Young}.) Then we consider the 
corresponding transposed branch, where we replace each representation $(\Lambda_+;\Lambda_-)$
by its transpose $(\Lambda_+^t;\Lambda_-^t)$.\footnote{The transpose $\Lambda^t$ is the Young tableaux
that is obtained from $\Lambda$ upon reflection along the diagonal.} Obviously  the transposed branch
shares the representation $(\tiny\yng(1);0)$ with the original branch; thus we should take it to start from
$(\tiny\yng(1,1);0)$ instead --- this then leads to the branch corresponding to the {\color{red} red} edges in Fig. \ref{Young}.
 Note that taking the transpose does not modify $B_-$ --- so the transposed
branch also has one representation at each level --- but  it changes the sign of both
$D_+$ and $D_-$, and hence the sign of $\Delta D$. Thus the change in conformal dimensions of the 
transposed representations are, apart from the `drift term' $(\lambda^2 B_- - 1)/N$, opposite
to those of the original representations, see eq.~(\ref{hdef}).

We now propose that these
two branches together are to be 
mapped to the first two families that come out of the perturbative analysis, i.e.\ the family 
(\ref{blockrecursion0}) that mixes 
with $(\tiny\yng(1);0)$, and the family (\ref{blockrecursion}) that mixes with the level $2$ state given by 
$\vec{f}^{\; (2)}_2$.
In particular, in the 't~Hooft limit the ratio $\Delta D / N$ (as well as the drift term $(\lambda^2 B_- - 1)/N$) become 
continuous variables, and hence the branches of minimal model representations lead to a continuum of 
perturbed conformal dimensions; this matches the continuum for $\rho$ we found above.
At finite $N$ and $k$, the two branches contain the same number of states as the two
families, and their eigenvalues are, except for the `drift term' $(\lambda^2 B_- - 1)/N$,
symmetrically distributed around $h$,  thereby matching the structure found in the perturbative analysis. 
(We shall  comment below on the origin of the `drift term' in the perturbative analysis.)

\smallskip

It should now be clear how to continue. While the above two branches account for all representations
up to level $2$, there are two more branches starting with edges linking level $2$ to level $3$. Again,
they can be completed into branches by picking at each level a representation with the 
second biggest (or smallest) value for $\Delta D$ --- these two branches can be chosen to be 
transposes of one another, and they correspond to the solutions of the perturbative analysis with 
$f_1=\vec{f}_2=0$ and $\vec{f}_3\neq 0$. (They are described by the {\color{blue} blue} and {\color{green} green}
edges in Fig. \ref{Young}.)
Furthermore, their eigenvalues will, again apart from the drift term,
be symmetrically distributed around $h$. 

Continuing in this manner, there will be $P(n)$ branches emerging at level $n$, and their conformal
weights will, apart from the drift term, be symmetrically distributed around $h$. 
This matches nicely the structure of the eigenvalues as computed in the perturbative
analysis. 

It remains to comment on the reason that the `drift term' $(\lambda^2 B_- - 1)/N$ in (\ref{hdef})
is invisible to the perturbation analysis of the
previous subsection. Recall that in the formula of the quadratic Casimir (\ref{casimir2}), the term proportional to 
$B$ is the leading term in the $\frac{1}{N}$ expansion. It  should therefore be considered as the `classical'
contribution, while the term proportional to $D$ corresponds 
to the first $\frac{1}{N}$ correction. Correspondingly, in the expansion of the conformal dimension (\ref{generalw}), 
although both $\frac{\lambda^2 B_{-} - 1}{2N}$ and $\frac{D_+- D_-}{2N} \, \lambda $ are of 
$\mathcal{O}(\frac{1}{N})$, in some sense only the first one  is `classical'.\footnote{
We remind the reader that the `$1$' in $\frac{\lambda^2 B_{-} - 1}{2N}$ comes from $B_{\Box}=1$.} Therefore 
in the large but finite $N$ case, we shouldn't merely truncate the infinite mixing matrix to a finite one; we should 
also shift the diagonal entries of the mixing matrix by the `classical' piece 
$(-4)\frac{\lambda^2 B_{-} - 1}{2N}$. Once this is done, the 
new eigenvalues are distributed symmetrically with respect to the shifted $\mathcal{O}(g^0)$ conformal 
dimension, thus explaining the `drift term' $\frac{\lambda^2 B_{-}-1}{N}$ in (\ref{hdef}). 
\medskip

We therefore regard this as good evidence for the assertion that the perturbation by $P$ corresponds to switching 
on the above $\frac{1}{N}$ corrections of the ${\cal W}_{N,k}$ minimal models in the 't~Hooft limit. 
We should also
mention that this identification fits with what we have seen explicitly for the special cases $\lambda=0$ and
$\lambda=1$ above. Indeed, for $\lambda=1$, it follows from (\ref{dgam}) that in order for $\delta\gamma^2=0$,
we need to take $\delta k=0$, see also (\ref{dkN}). But then, it follows from (\ref{dlam}) and
(\ref{dc}) that both $\delta \lambda = \delta c =0$, i.e.\ that the perturbation does not change anything,
in nice agreement with what we saw in Section~\ref{sec:pertmu1}. From the point of view of the
perturbative analysis, at $\lambda=1$ where $N\rightarrow \infty$ at finite $k$, it is no longer appropriate to make a 
$\frac{1}{N}$ expansion, but we should rather perform a $\frac{1}{k}$ expansion, replacing 
$\frac{1}{N}$ by $\frac{1}{k}$ as in (\ref{Nkrep}). But then (\ref{Pij}) vanishes
at $\lambda=1$, i.e.\ the spectrum is not perturbed at all.

On the other hand, for $\lambda=0$, $\delta \lambda=0$ automatically from (\ref{dlam}) and $\delta c = \delta N$ from (\ref{dc}). In this case the condition $\delta \gamma^2=0$ from (\ref{dgam}) does not impose any restriction on $\delta k$ and $\delta N$,
and thus $\delta c = \delta N$ will be non-zero. Note, however, that the coefficient of the $\Delta D$-term 
in  (\ref{hdef}) vanishes for $\lambda=0$. This is reflected, in the context of the 
perturbation computation, by the fact that the analysis breaks down for $\lambda=0$ as the $B_n$ do not have 
maximal rank any longer. (For instance, the rank of $B_3$ is ${\rm rk}(B_3)=3<4$, at $\lambda=0$.) 
This reduction in the rank of $B_n$ at $\lambda=0$  is a sign that 
the scalar-like states do not `mix' strongly enough to break the degeneracy of their conformal dimensions.

\section{Conclusions}\label{sec:conc}

In this paper we have studied the behaviour of the ${\cal W}_{\infty}[\lambda]$ 
theories under perturbations that preserve the symmetry algebra to first order. In particular,
we have found that the minimal models possess such a perturbing field in the 't~Hooft limit, and we 
have shown that it corresponds to switching on the $\frac{1}{N}$ corrections, while
keeping $\lambda$ fixed at first order. Since the theory at finite $N$ involves necessarily
the light states (that may be taken to decouple in the 't~Hooft limit \cite{Gaberdiel:2011zw}), it is not surprising
that these states, as well as their descendants, play an important part in the analysis. 
The perturbative analysis is technically rather demanding since the strict 't~Hooft limit
is a degenerate point where infinitely many states --- the analogues of the light states at 
$h=\frac{1}{2}(1+\lambda)$ --- have the same conformal 
dimension and hence can mix in perturbation theory. However, as we have seen, the
structure of the theory is sufficiently rigid so as to allow one to understand at least some of the
qualitative features.

We should mention that the 2d case we have considered here differs qualitatively from
what happens in higher dimensions. In particular, it was shown in \cite{Maldacena:2011jn,Maldacena:2012sf} 
that for $d\geq 3$, the $\frac{1}{N}$ corrections necessarily break the higher spin symmetry to first order. 
This is to be contrasted with what we have found here, namely that there are perturbing 
fields that preserve the symmetry at least to first order.

Perturbations of these ${\cal W}_{\infty}[\lambda]$ algebras, not necessarily preserving the higher spin
symmetry, will also be important in order to connect suitable generalisations of these theories
to string theory. In particular, it would be interesting to understand what the effect of the
exactly marginal perturbing field of the large ${\cal N}=4$ higher spin theory of 
\cite{Gaberdiel:2013vva} is. It should be possible to study this question with similar techniques as 
those used in the present paper.

\section*{Acknowledgements}

We thank Stefan Fredenhagen, Rajesh Gopakumar and Igor Klebanov for useful discussions,
Maximilian Kelm and Carl Vollenweider for providing us access to their unpublished results
\cite{MC}, and Maximilian Kelm for explanations and discussions.
The work of MRG and KJ is supported in parts by the Swiss National Science Foundation.
Part of this work was done while we were visiting the GGI during the programme on 
`Higher Spins, Strings and Duality'. 
We thank the Galileo Galilei Institute for Theoretical Physics for the hospitality and the INFN for partial 
support. MRG and WL thank IPMU, and WL thanks KEK and the Benasque string workshop for their hospitality during
various stages of this work.

\appendix

\section{The free fermion theory at $\lambda=0$}\label{app:OPE}

In this appendix we collect some of the OPEs we have calculated for the free fermion 
theory. Using Wick's theorem it is not difficult to work out the singular part of the OPEs of these currents. In particular, we find that
\begin{eqnarray}
J(z_1) \, J(z_2) & \sim & \frac{N}{(z_1-z_2)^2}  \label{JJ}\\[4pt]
T(z_1) \, T(z_2) & \sim & \frac{N/2}{(z_1-z_2)^4} + \frac{2 \, T(z_2)}{(z_1-z_2)^2} + \frac{T'(z_2) }{(z_1-z_2)} \\[4pt]
T(z_1)\, J(z_2) & \sim & \frac{J(z_2)}{(z_1-z_2)^2} + \frac{J'(z_2)}{(z_1-z_2)} \\[4pt]
J(z_1)\, T(z_2) & \sim & \frac{J(z_2)}{(z_1-z_2)^2} \ ,  \label{JT} \\[4pt]
J(z_1) W(z_2) & \sim & \frac{2}{(z_1-z_2)^2} \, T(z_2)  \label{JW} \\[4pt]
T(z_1) W(z_2) & \sim & \frac{1}{(z_1-z_2)^4} \, J(z_2) + \frac{3}{(z_1-z_2)^2} W(z_2) + \frac{1}{(z_1-z_2)} W'(z_2)  \ , \label{TW} \\[4pt]
W(z_1) W(z_2) &\sim & \frac{2N/3}{(z_1-z_2)^6} + \frac{4 T(z_2)}{(z_1-z_2)^4} + \frac{2 T'(z_2)}{(z_1-z_2)^3}
+ \frac{ 4 U(z_2) + \frac{3}{5} T''(z_2)}{(z_1-z_2)^2} 
 \nonumber \\
& & \ + \frac{ 2 U'(z_2) + \frac{2}{15} T'''(z_2)}{(z_1-z_2)} \ , \label{WW}
\end{eqnarray}
where $\sim$ always denotes the singular part of the operator product expansion.

\section{The free boson theory at $\lambda=1$}\label{app:bos}

The first few OPEs of the free boson higher spin fields defined in eq.~(\ref{Wbos}) are explicitly given by
($z_{12}\equiv z_1 - z_2$)
\begin{eqnarray}
T(z_1) W^3(z_2) &\sim & \frac{3}{z_{12}^2} W^3(z_2) + \frac{1}{z_{12}} \partial W^3(z_2)  \label{bos23} \\
T(z_1) W^4(z_2) &\sim & 
\frac{192}{5} \frac{T(z_2)}{z_{12}^4} + \frac{4}{z_{12}^2} W^4(z_2) + \frac{1}{z_{12}} \partial W^4(z_2) \\[4pt]
W^3(z_1) W^3(z_2) &\sim & 
48 \left[ \frac{T}{z_{12}^4} + \frac{1}{2}\frac{\partial T}{z_{12}^3} + \frac{3}{20}\frac{\partial^2 T}{z_{12}^2} + \frac{1}{30}\frac{\partial^3 T}{z_{12}} \right] \nonumber \\[2pt]
&& + \frac{4}{z_{12}^2} W^4 (z_2) + \frac{2}{z_{12}} \partial W^4(z_2) + 16 \frac{k} 
{z_{12}^6} \\[4pt]
W^3(z_1) W^4(z_2) &\sim & 
\frac{768}{5} \left[ \frac{W^3}{z_{12}^4} + \frac{1}{3}\frac{\partial W^3}{z_{12}^3} + \frac{1}{14}\frac{\partial^2 W^3}{z_{12}^2} + \frac{1}{84}\frac{\partial^3 W^3}{z_{12}} \right] \nonumber \\[2pt]
&& + \frac{5}{z_{12}^2} W^5 (z_2) + \frac{2}{z_{12}} \partial W^5(z_2) \\[4pt]
W^4(z_1) W^4(z_2) &\sim & 
\frac{12288}{5} \left[ \frac{T}{z_{12}^6} + \frac{1}{2} \frac{\partial T}{z_{12}^5} 
+ \frac{3}{20} \frac{\partial^2 T}{z_{12}^4}
+ \frac{1}{30} \frac{\partial^3 T}{z_{12}^3} + \frac{1}{168} \frac{\partial^4 T}{z_{12}^2} 
+ \frac{1}{1120} \frac{\partial^5 T}{z_{12}} \right] \nonumber \\[2pt]
&& + 
\frac{1728}{5} \left[ \frac{W^4}{z_{12}^4} + \frac{1}{2} \frac{\partial W^4}{z_{12}^3} 
+ \frac{5}{36} \frac{\partial^2 W^4}{z_{12}^2} + \frac{1}{36} \frac{\partial^3 W^4}{z_{12}} \right] \nonumber \\[2pt]
&& + \frac{6}{z_{12}^2} W^6(z_2) + \frac{3}{z_{12}} \partial W^6(z_2) + \frac{3072}{5} \frac{k} 
{z_{12}^8} \ . \label{bos44}
\end{eqnarray}

\section{Explicit mixing matrices}\label{sec:mixmat}

Using the notation of (\ref{Jacobi}), the first few explicit expressions for the $B_n$ are as follows
\begin{itemize}
\item $\frac{N}{16 \pi g} \, B_1 \, :$
\begin{table}[!h]
\centering
\begin{tabular}{| c | c  c | }
\hline
& $(\tiny\yng(1,1);\tiny\yng(1))$ & $(\tiny\yng(2);\tiny\yng(1))$   \cr
\hline
$(\tiny\yng(1);0)$ & $\frac{1-\lambda}{\sqrt{2}}$  & $\frac{1+\lambda}{\sqrt{2}}$ \cr 
\hline
\end{tabular}
\end{table}
\item $\frac{N}{16 \pi g} \, B_2 \, :$
\begin{table}[!h]
\centering
\begin{tabular}{|c | c  c  c  c |}
\hline
& $(\tiny\yng(1,1,1);\tiny\yng(1,1))$ & $(\tiny\yng(2,1);\tiny\yng(2))$ & $(\tiny\yng(2,1);\tiny\yng(1,1))$ &$(\tiny\yng(3);\tiny\yng(2))$  \cr
\hline
$(\tiny\yng(1,1);\tiny\yng(1))$ & $\sqrt{\frac{2}{3}} (1-\lambda) $  & 0 & $\frac{2+\lambda}{\sqrt{3}} $&0\cr 
$(\tiny\yng(2);\tiny\yng(1))$ & 0& $\frac{2-\lambda}{\sqrt{3}} $ &0& $\sqrt{\frac{2}{3}} (1+\lambda) $ \cr 
\hline
\end{tabular}
\end{table}
\item $\frac{N}{16 \pi g} \, B_3 \, :$ 
\begin{table}[!h]
\centering
\begin{tabular}{|c | c c c c c c c | }
\hline
& $(\tiny\yng(2,2);\tiny\yng(2,1))$ & $(\tiny\yng(2,1,1);\tiny\yng(2,1))$ & $(\tiny\yng(3,1);\tiny\yng(2,1))$ &$(\tiny\yng(1,1,1,1);\tiny\yng(1,1,1))$  &$(\tiny\yng(2,1,1);\tiny\yng(1,1,1))$  &$(\tiny\yng(3,1);\tiny\yng(3))$ &$(\tiny\yng(4);\tiny\yng(3))$ \cr
\hline
$(\tiny\yng(1,1,1);\tiny\yng(1,1))$  & $0$  & $0$ & $0$ &  $\frac{\sqrt{3}(-1+\lambda)}{2}$ &$ -\frac{3+\lambda}{2}$ &0&0\cr 
$(\tiny\yng(2,1);\tiny\yng(2))$&$\frac{\sqrt{3}(1-\lambda)}{2\sqrt{2}}$ & $\frac{3-\lambda}{4}$ &$\frac{3(1+\lambda)}{4}$& 0 &0&0&0\cr 
$(\tiny\yng(2,1);\tiny\yng(1,1))$&$-\frac{\sqrt{3}(1+\lambda)}{2\sqrt{2}}$& $\frac{3(-1+\lambda)}{4}$ &$-\frac{3+\lambda}{4}$& 0 &0&0&0\cr 
$(\tiny\yng(3);\tiny\yng(2))$& 0& 0 &0& 0 &0& $ \frac{3-\lambda}{2}$ &$\frac{\sqrt{3}(1+\lambda)}{2}$ \cr 
\hline
\end{tabular}
\end{table}
\end{itemize}

\section{The eigenproblem of a scalar Jacobi matrix}\label{sec:mixmatscalar}

Let us consider the perturbative solution corresponding to $f_1=1$, i.e.\ the solution that involves
the original scalar field ${\cal O}\equiv (\tiny\yng(1),0)$ itself. As was argued in section~\ref{sec:eigenv}, the
perturbative analysis will mix to this state one specific linear combination of scalar-like states at each level; thus we can 
consider the truncated problem, where instead of the full matrix $M$ we just consider the eigenvalue 
problem 
\begin{equation}
J\cdot\vec{F}=\rho\, \vec{F}  \qquad \qquad \textrm{with} \quad\vec{F}=(f_1,f_2, \ldots)
\end{equation}
for the `scalar Jacobi matrix' 
\begin{equation}\label{Jacobiscal}
 J(\{a_{n}\}, \{b_n\})= \begin{bmatrix}
    a_1     & b_{1} & 0 & 0 &0&\ldots \\
    b_{1} & a_2     & b_2 &0 &0&\ldots \\
0& b_{2} & a_3     & b_{3} &0 &\ldots \\

   0 & 0 & \ddots& \ddots& \ddots & 
  \end{bmatrix} \ ,
\end{equation}
where now all $a_n$ and $b_n$ are numbers. This can be solved  recursively; since the value of $f_1$ 
only affects the overall normalisation, we can start by setting $f_1=1$. Then given any
$\rho\in \mathbb{C}$, we solve for $f_n(\rho)$ as 
\begin{eqnarray}\label{recursion}
&&f_1=1 \nonumber\\
a_1 f_1+b_1 f_2=\rho\, f_1 &\rightarrow& f_2=\frac{1}{b_1}(\rho-a_1)f_1\nonumber\\
&\dots&\\
b_{n-1}f_{n-1}+a_n f_n+b_{n}f_{n+1}= \rho\, f_n &\rightarrow& f_{n+1}=\frac{1}{b_n}\bigl[(\rho-a_n)f_{n}-b_{n-1}f_{n-1}\bigr]
\nonumber\\
&\dots&\nonumber
\end{eqnarray}
Note that, by construction, $f_n(\rho)$ is a polynomial of degree $n-1$ in $\rho$. If the Jacobi matrix is strictly infinite, this is the 
most general solution.
\medskip

If the Jacobi matrix truncates to some finite matrix, say to a matrix of size $K\times K$ --- this will be the case
for finite $N$ and $k$ ---  then the analysis
can be performed similarly. We simply extend the Jacobi matrix to an infinite matrix by choosing the $a_n$ and 
$b_n$ for $n> K$ arbitrarily. Then we can construct recursive eigenvalues as above. Up to level $K$
the solution agrees with the solution to the actual eigenvalue problem we are interested in, so all we have to 
require is that the analysis terminates at level $K$, i.e.\ that $f_{K+1}(\rho)=0$. Since $f_n(\rho)$ is a polynomial of 
degree $n-1$ in $\rho$, this gives rise to $K$ different solutions, as expected. Note that if all $a_i$, $i=1,\ldots ,K$ 
agree, $a_i\equiv a$, then $f_n(\rho)$ is an even (odd) polynomial of $\rho-a$ if $n$ is odd (even); in this case the eigenvalues
will be symmetrically distributed around $\rho=a$. If the diagonal entries exhibit some `drift', we expect that 
also the eigenvalues will become symmetrically distributed w.r.t.\ some shifted mean.

\begin{singlespace}

\end{singlespace}

\end{document}